\input psfig
\documentclass[preprint]{aastex}

\newcommand{\Ms}{M$_{\odot}$}
\newcommand{\mlow}{M$_{\rm{low}}$}
\newcommand{\mup}{M$_{\rm{up}}$}
\newcommand{\Zs}{Z$_{\odot}$}
\newcommand{\cB}{MS~1512-cB58}

\newcommand{\kms}{km~s$^{-1}$}

\newcommand{\civ}{C~IV $\lambda$1550}
\newcommand{\siiv}{Si~IV $\lambda$1400}
\newcommand{\nv}{N~V $\lambda$1240}

\slugcomment{To appear in The Astrophysical Journal}

\shorttitle{Metal-Poor Star-Forming Galaxies}
\shortauthors{Leitherer et al.}

\begin{document}
\title{Ultraviolet Line Spectra of Metal-Poor Star-Forming Galaxies}

\author{Claus Leitherer}
\affil{Space Telescope Science Institute\altaffilmark{1}, 
       3700 San Martin Drive,
       Baltimore, MD 21218\\
       leitherer@stsci.edu}

\author{Jo\~ao R. S. Le\~ao}
\affil{Depto de F\'{\i}sica, CFM -- UFSC, CP~478, 
     88040-900 Florian\'opolis, SC, Brazil\\
      joao@fsc.ufsc.br}
      
\author{Timothy M. Heckman}
\affil{Johns Hopkins University, Dept. of Physics \& Astronomy,
       Baltimore, MD 21218\\
        heckman@pha.jhu.edu}

\author{Daniel J. Lennon}
\affil{Isaac Newton Group, Apartado 321, 38700 Santa Cruz de La Palma,
   Canary Islands, Spain\\
    djl@ing.iac.es}

\author{Max Pettini}
\affil{Inst. of Astronomy, Univ. of Cambridge, Madingley Road,
     Cambridge CB3 0HA, UK\\
       pettini@ast.cam.ac.uk}

\and

\author{Carmelle Robert}
\affil{D\'ept. de Physique, Universit\'e de Laval, Qu\'ebec, QC, G1K 7P4,
    Canada\\
    carobert@phy.ulaval.ca}

\altaffiltext{1}{Operated by AURA, Inc., under NASA contract NAS5-26555}


\begin{abstract}
We present synthetic ultraviolet spectra of metal-poor star-forming galaxies 
which were calculated with the Starburst99 package. A new spectral library 
was generated from HST observations of O stars in the Large and Small
Magellanic Clouds. The corresponding mean metallicity of the synthetic spectra
is approximately $\frac{1}{4}$~\Zs. The spectra have a resolution of 1~\AA\ 
and cover the spectral range 1200~--~1600~\AA. A set of model spectra was
calculated for a standard initial mass function and star-formation history
and is compared to synthetic spectra at solar metallicity. We find 
that the spectral lines are generally weaker at lower metallicity, as 
expected from the lower elemental abundances. Stellar-wind lines, however, 
show a more complex behavior: the metallicity dependence  of the 
ionization balance can be important in trace ions, like N$^{4+}$ and 
Si$^{3+}$. Therefore the strength of \nv\ and \siiv\ does not scale 
monotonically with metallicity. We compare our new models to ultraviolet
spectra of NGC~5253 and \cB, two star-forming galaxies with 
$\frac{1}{4}$~solar metallicity at low and high redshift, respectively. The
new library provides significantly better fits to the observations than earlier
models using the \Zs\ library. We discuss the potential of utilizing stellar
photospheric and wind lines to estimate the chemical composition of 
star-forming galaxies. The new metal-poor synthetic spectra are available
via the Starburst99 website.

\end{abstract}


\section{Introduction: star-forming galaxies in the ultraviolet}

The ultraviolet (UV) to near-infrared (IR) line spectra of galaxies with
current star formation exhibit a pronounced dichotomy: the UV region is
dominated by strong stellar and interstellar {\em absorption} lines whereas
in the optical and near-IR
such lines (if present at all) are hidden by strong interstellar
{\em emission} lines (see Fig.~13 of Kinney et al. 1996). Therefore the
UV is the wavelength region of choice for the direct detection of stellar
features in star-forming galaxies. The most extensive documentation of
the UV spectral morphology of star-forming galaxies was published by Kinney
et al. (1993), who compiled an atlas of all usable spectra in the IUE archive.
More recently, HST's superior sensitivity and spatial as well as spectral 
resolution have made it possible to study a number of galaxies in much greater
detail (Leitherer 2000).

The UV line spectra of local star-forming galaxies have become an important
tool for the interpretation of the restframe UV of star-forming galaxies
at high redshift (Conti, Leitherer, \& Vacca 1996). The similarity of the
spectra at low and high redshift suggests similar stellar content
(Ebbels et al. 1996; Steidel et al. 1996; Yee et al. 1996; Dey et al. 1997;
Lowenthal et al. 1997; Pettini et al. 2000).

The underlying physics for the spectral appearance is rooted in the atmospheric
conditions of hot, massive stars. Stars with masses upward of $\sim$10~\Ms\
and temperatures higher than $\sim$25,000~K provide the UV light in
star-forming galaxies. Such stars have strong stellar winds with velocities
up to 3000~\kms, leading to strong blueshifted absorption lines or
even P~Cygni-type profiles in the UV (Walborn, Nichols-Bohlin, \& Panek
1985). The strongest lines are typically \nv, \siiv, and \civ. Since the
properties of hot, massive stars change rapidly over cosmological time-scales,
their line profiles in galaxy spectra do so as well and are useful as tracers
of the massive stellar population. The promise of utilizing stellar 
absorption lines of \siiv\ and \civ\ to infer the massive-star content was
first recognized by Sekiguchi \& Anderson (1987a, b) who related their
{\em equivalent widths} to the stellar initial mass function (IMF). The
relative strengths of these two lines can be used as a an indicator 
of massive stars (e.g., Mas-Hesse \& Kunth 1999), as well as of the kinematic
conditions of the interstellar medium (Heckman et al. 1998).

The earlier IUE work using  this diagnostic tool was limited to equivalent
width determinations in galaxy spectra. With HST, UV {\em line-profile} 
studies are feasible, and the full information of the profile shapes can
be taken into account. This becomes particularly crucial for galaxies with
strong interstellar lines whose interstellar and stellar-wind components are
often severely blended. Semi-empirical UV spectral templates 
at 0.75~\AA\ resolution were computed from evolutionary
synthesis models by Robert, Leitherer, \& Heckman (1993), Leitherer, Robert,
\& Heckman (1995), and de~Mello, Leitherer, \& Heckman (2000). These models
linked a stellar UV library to the Starburst99 code (Leitherer et al. 1999)
to produce a grid of UV line spectra for a standard set of stellar parameters.
In many cases the calculated spectra are in rather good agreement with
observed galaxy spectra (e.g., Johnson et al. 1999; Pettini et al. 2000;
Tremonti et al. 2000).

In our previous modeling, the spectral library used
for the comparison with the galaxy spectra was built from hot stars with
approximately solar or slightly sub-solar chemical composition although
 the  galaxies were known to be more metal-poor. Until recently, no 
 library of metal-poor stars was available since IUE's high-dispersion
 mode is required for adequate spectral resolution, and therefore only 
 nearby bright
 OB stars could be observed. The stars of choice would be in the Magellanic
 Clouds where OB stars have metallicities lower by factors of 3~--~10 than
 their Galactic counterparts in the vicinity of the Sun (Garnett 1999).  
 Observations of hot stars in the Large (LMC) and Small (SMC) Magellanic Clouds
have been
 accumulated with HST's UV spectrographs over several cycles in the past.
These data are suitable for the construction of a low-metallicity UV library
of hot stars as discussed in the preliminary work of Robert (1999a).

The purpose of this paper is to report the availability of a UV spectral 
library of LMC and SMC stars and to discuss its application to the study of
the
stellar content of star-forming galaxies with metallicities similar to those
of the Magellanic Clouds. In Section~2 we describe the data set and the
 reduction procedure.
The generation of the library and its properties are discussed in Section~3.
In Section~4, a series of synthetic low-metallicity spectra is presented. The
synthetic spectra are compared to observations of NGC~5253 and \cB, two
metal-poor galaxies at low and high redshift, respectively (Section~5). 
Conclusions are given in Section~6.

\section{Observations and data reduction}

\subsection{Sample definition}

The library spectra were obtained with four HST General Observer (GO) programs
between Cycles~1 and 7: 2233 (PI Kudritzki), 4110 (PI Kudritzki; carry-over
program of 2233), 5444 (PI Robert), and 7437 (PI Lennon). Programs 5444 and 
7437 had the
specific science goals of creating a UV spectral library of LMC/SMC stars.
The goal of programs 2233 and 4110 was to obtain spectra of hot stars
 for an atmospheric analysis; since the data are perfectly
adequate for our intended library, they were included as well. We visually
inspected the spectra of all targets in the four programs
using the HST archive preview tool. Failed observations and spectra with very
low S/N were discarded. 59 useful spectra were selected for a total of 53
stars which are listed in Table~1.

The library stars in Table~1 are arranged by advancing spectral type. The
sources for the identifiers are as follows: NGC~346 is an H~II region in
the SMC and the star numbers are in the nomenclature by Massey, Parker,
\& Garmany (1989) and Walborn et al. 
(1995). Note that \#435, \#355, \#342, and \#324
 of Massey et al. (1989) are \#1,
\#3, \#4, and \#6 of Walborn et al. (1995), respectively. ``Sk'' stands for
Sanduleak and denotes entries from Sanduleak's (1969) atlas of the LMC. 
``AV'' entries are from the catalog of early-type SMC stars by Azzopardi
\& Vigneau (1975, 1982). The ``BI'' label was introduced by Crampton (1979) and
refers to the LMC OB-star catalog of Brunet et al. (1975). The identifiers
in Table~1 are the same as used in the original HST programs in order to
facilitate matching the spectra presented here with the original data ---
with these exceptions: \#355 and \#324 are called \#3 and \#6 in program
4110; all ``Sk'' stars in program 5444 are called ``NS'' (for Nick Sanduleak),
and the declination zone and running index are inverted (e.g., 
Sk--69$^\circ$124 becomes NS124-69).

The stellar spectral types (col.~3 of Table~1) are based on the classification
work of Conti, Garmany, \& Massey (1986), Garmany, Conti, \& Massey (1987),
Walborn et al. (1995), and Walborn et al. (2000). Spectral coverage is 
complete from O3 through B0 if the whole LMC and SMC sample is considered,
but significant spectral gaps are present for each galaxy sample individually.
Therefore we decided to build a mean library of LMC and SMC stars, with a
corresponding mean metallicity of about $\frac{1}{4}$~\Zs. We searched the
HST archive for suitable spectra of stars with later spectral types but none
were found. The color excess $E(B-V)$ in col.~4 of Table~1 was calculated
from the observed and the intrinsic $(B-V)$. The photometry was taken from
the same sources as the spectral types. The intrinsic color was assumed to
be $(B-V) = -0.31$, independent of spectral type. The mean $E(B-V)$ of all
53 stars is 0.11.

\subsection{STIS data}

A description of the technical aspects of the STIS data set was given by
Walborn et al. (2000). The data were taken with the STIS FUV-MAMA detector in
the E140M mode. The $0.2'' \times 0.2''$ entrance aperture was used. This
instrument set-up provides a 1-pixel resolving power of $R = 92,000$ over 
a wavelength range of 1150 to 1700~\AA. 

The data were retrieved from the HST archive and recalibrated with the
IRAF/STSDAS software at STScI. The standard ``calstis'' calibration pipeline 
steps were followed, except for the 1-dimensional spectrum extraction. The
standard pipeline does not adequately correct for the underlying background
due to scattered light. Therefore a proto-type version of the ``x1d'' package
was used which corrects for scattered light with the algorithm of Howk \&
Sembach (2000). Tests with the black Milky Way Lyman-$\alpha$ absorption
trough in the spectra demonstrated that this algorithm successfully accounted
for the scattered light. After the individual orders were extracted, they were
merged into a single spectrum with a beta-version of the ``splice'' 
application within the ``calstis'' package. This application merges the
orders with an optimized solution for the overlap region.

Following the basic reduction, several additional reduction steps were 
applied. First we co-added multiple exposures of the same stars. Then
we normalized the continuum to unity after division by a
Chebyshev polynomial fitted to line-free sections of the spectrum.
The continuum normalization  process removed a flux depression
 between 1435
and 1440~\AA\ due to shadowing by the STIS MAMA repeller wire. The 
effect on the spectrum is a lower S/N in this wavelength region. 
Next, a
velocity offset of --150~\kms\ was added to account for the recession velocity
of the SMC (all LMC stars were observed with FOS). 
Finally, the spectra were resampled and truncated to match the
spectral resolution and wavelength coverage of the spectra obtained with the
FOS. The sampling and spectral resolution were decreased to 
0.75~\AA\ per pixel. The starting  wavelength was set to
1205.5~\AA, which is the same as that of
the \Zs\ library. The spectra were truncated at
1601~\AA, since the FOS data terminate at that wavelength.

The final reduced spectra are essentially identical to those of Walborn
et al. (2000), which were processed with an independent algorithm.

\subsection{FOS Snapshot data}

The FOS Snapshot spectra of program 5444 were taken with the blue FOS
detector through the $1.2'' \times 3.7''$ entrance aperture. The target
acquisition mode was ``point-and-shoot'', i.e. the spectra were taken 
without additional centering after the initial guide-star acquisition. Due 
to the large aperture used and the expected pointing error of $1'' - 2''$
most (but not all) observations succeeded and well-exposed spectrograms
are in the HST archive. Each star was observed with two gratings, G130H and
G190H, resulting in a wavelength coverage of 1140~--~1605~\AA\ and
1573~--~2330~\AA, respectively.  The nominal spectral resolution for a 
point source observed with G130H is 1.0~\AA.
Since the STIS spectral coverage ends at
1700~\AA, the same maximum wavelength should apply to the FOS data as well.
We decided not to include the G190H spectra in the library because (i) the
extra $\sim$100~\AA\ would not add other strong stellar lines in a
galaxy spectrum, and (ii) the 1600~--~1700~\AA\ region is at the 
sensitivity boundaries for both gratings, with low count numbers and large
calibration uncertainties. Unfortunately, the (generally weak) 
He~II $\lambda$1640 line is longward of this cut-off and cannot be 
synthesized.

The retrieved spectra were processed with the ``calfos'' pipeline within
IRAF/STSDAS. The processed spectra were not significantly different from
the calibrated spectra in the archive. Several spectra showed spurious 
emission lines around 1430 and 1510~\AA. In all cases they
turned out to be artifacts due to intermittent diodes. The features were
successfully removed by interpolation between the adjacent continuum.
 The subsequent data
processing was similar to that applied to the STIS data. Multiple exposures
were co-added, the continuum was normalized to unity, and radial velocity
corrections of --150~\kms\ and --270~\kms\ were added to the SMC and LMC
data, respectively. Most spectra showed significant wavelength offsets even
after the radial velocity correction. This was expected since the 
point-and-shoot target acquisition can place the star anywhere inside
(and sometimes outside) the $1.2'' \times 3.7''$ aperture, which extends
over four science diodes. Therefore wavelength shifts of $\pm2$~\AA\ can
occur. The shifts were corrected by establishing a wavelength zero point
with  strong interstellar lines
in the STIS spectra, and then determining the offset in the FOS data.

The final step was to resample the spectra to 0.75~\AA\ per pixel and to
truncate them at the same boundaries as the STIS data. A value of
0.75~\AA\ is slightly higher but still close to the nominal spectral 
resolution of 1.0~\AA.

\subsection{FOS GO data}

The observations from GO programs 2233 and 4110 were performed prior to
the installation of Costar. As in the Snapshot program, G130H and
G190H spectra were taken. Since the data were obtained in the aberrated
beam, the narrow $0.25'' \times 2.0''$ entrance slit was used to retain
the full nominal spectral resolution. A full target acquisition, including
peak-ups, was required to center the star.

As we did with the Snapshot data, only the G130H spectra were considered.
All reduction steps were analogous to the Snapshot processing. Although these
spectra did not suffer from the large wavelength offsets due to a missing
target acquisition, we nevertheless found that the standard FOS wavelength
calibration could be improved by comparing them to the STIS data. Therefore
we determined and applied the wavelength offsets in the same way as we did
for the Snapshot data. The final product was a wavelength calibrated, 
normalized spectrum from 1205.5 to 1601~\AA\ at 0.75~\AA\ sampling for each
star.

We coadded the spectra of six stars with data  
from more than one program: NGC~346\#355, NGC~346\#324,
Sk--70$^\circ$69, AV~75, AV~15, and AV~47. This then resulted in a total
of 53 unique stellar templates with typical S/N of about 30 (as determined
mostly by the FOS data; the STIS spectra have higher S/N).

\section{Generation of the library}

The low-metallicity library was built with the same rules we used for the
solar-metallicity library in Starburst99 (Robert et al. 1993; de~Mello et al.
2000). We averaged the spectra of stars with the same spectral type in order
to increase the S/N and to minimize the impact of spectral misclassifications.
Although only stars with classifications based on slit spectra are considered,
the spectral types of some stars are less certain due to the quality
of the optical classification spectra. The format
of the library is a two-dimensional grid with 15 temperature classes between
O3, O3.5, O4, O4.5,.........B0 and five luminosity classes of V, IV, III, II,
and I. This grid is dictated by the interpolation scheme in Starburst99 and
does not directly correspond to the MK classification grid of O stars.
The spectral types in Table~1 were sufficient to populate about 40\%
of this grid. Missing entries (mostly stars with luminosity classes IV and
II) were filled by two-dimensional interpolation between adjacent temperature
and luminosity classes. 

The conversion from spectral type to temperature and luminosity is done with
the prescription of Schmidt-Kaler (1982). This relation is somewhat
different for O stars than that of Vacca, Garmany, \& Shull (1996). As a 
test, we selected the subset of 15 SMC O stars discussed by Walborn et al.
(2000) and placed them on a Hertzsprung-Russell diagram both using our
calibration and that of Vacca et al. The mean temperature and  log bolometric
luminosity differences (Schmidt-Kaler -- Vacca) are $-2100 \pm 800$~K and
$0.03 \pm 0.12$, respectively. The average
temperature shift is smaller than 5\%
and corresponds to less than one spectral sub-type. These are the uncertainties
that should be kept in mind when reconstructing a Hertzsprung-Russell
diagram with our method. Clearly the stellar positions on such a 
diagram predicted by Starburst99 have larger errors than those derived
via a dedicated atmospheric analysis. In addition, if the temperature
and luminosity scales of Schmidt-Kaler needed significant revision, 
the assigniment of library spectra to positions in the Hertzsprung-Russell
diagram would be modified as well. In our opinion, there is no clear
evidence for substantial flaws in either the Schmidt-Kaler or the 
Vacca et al. scales, and the differences simply reflect current
uncertainties in hot-star models. 

In Figs.~\ref{o4v}~--~\ref{o9i} we show examples of four spectral groups with
types O4~V, O9~V, O4~I, and O9~I. The four groups are representative for the
temperature- and luminosity-behavior of the strongest strategic lines of
\nv, \siiv, and \civ. A discussion of the properties of these lines at solar
metallicity was given in Leitherer et al. (1995). The corresponding spectral
groups at \Zs\ are included in Figs.~\ref{o4v}~--~\ref{o9i} for comparison.
As expected, the spectral features are generally (but not always) weaker at
lower metallicity (see the discussion by Robert 1996, 1999b and
Plante 1998). Photospheric lines
are uniformly weaker in the low-metallicity library  due to the lower opacity
in the Magellanic Clouds (Haser et al. 1998). Stellar-wind lines are more
complex, as both metallicity and wind properties determine their opacities.
Under most wind conditions N$^{4+}$, Si$^{3+}$, and C$^{3+}$ are trace ions
(Lamers et al. 1999) so that slight changes in the stellar temperature
can increase or decrease the line strengths significantly.

Early-O main-sequence stars (Fig.~\ref{o4v}) display wind effects in \nv\ and
\civ, but not in \siiv. This was first noted by Walborn \& Panek (1984) and
explained as an ionization effect of Si$^{3+}$ in main-sequence
stars with less dense winds by Drew (1990) and Pauldrach et al. (1990). The
Si~IV doublet absorption in {\em any} O main-sequence star is almost 
exclusively interstellar. \nv\ has no significant metallicity dependence in
the O4~V group. (The N~V profile shortward of 1230~\AA\ is blended with 
interstellar and 
geocoronal Lyman-$\alpha$.) The behavior of the \nv\ line is a prime example
of the counteracting effects of chemical composition and stellar temperature
on the N$^{4+}$ column density. The dominant ionization stage in O-star
winds is N$^{3+}$ and the mean ionization fraction of N$^{4+}$ increases
monotonically with stellar temperature (Lamers et al. 1999). Therefore, to
first order, the N$^{4+}$ becomes independent of the stellar mass loss because
the ionization fraction and the total nitrogen column density
($ = \sum N^i$) have the opposite dependence on $\dot{M}$. Full non-LTE
model atmosphere calculations by Kudritzki (1998; his Fig.~7) agree with this
plausibility argument: The \nv\ line is stronger in models at 
$Z= \frac{1}{5}$~\Zs\ than at \Zs.
In contrast to \nv, \civ\ is much weaker and the 
absorption is less blueshifted at lower metallicity, a direct result of the
lower wind density and velocity in the Magellanic Clouds. Ionization effects
are less important for C$^{3+}$ so that the column density of the C$^{3+}$
ion scales with the total column density of carbon.

The same trends are present in late-O main-sequence stars (Fig.~\ref{o9v}). The
wind effects are dramatically reduced for O9~V stars due to their much
lower luminosity. The only indication for a stellar wind at low metallicity 
is the asymmetric \civ\ profile. At solar metallicity, this line becomes
blueshifted due to the increased wind opacity. 

Stellar mass loss increases with luminosity as $\dot{M} \propto L^{1.5}$
(Garmany \& Conti 1984), and wind velocities are proportional to the
surface escape velocity: $v_{\infty} \propto v_{\rm{esc}}$ (Lamers, Snow,
\& Lindholm 1995). Therefore the wind densities increase by an order
of magnitude from luminosity class V to I for the same spectral type.
\nv, \siiv, and \civ\ have strong P~Cygni profiles over most of the O 
supergiant spectral
sequence due to the enhanced wind opacity. Both early-O (Fig.~\ref{o4i})
and late-O (Fig.~\ref{o9i}) supergiants in the Magellanic Clouds and in the
Galaxy have similar profiles, suggesting that metallicity becomes less 
important for more saturated line profiles, in comparison with the (mostly)
optically thin profiles of main-sequence stars. There is, however, a clear
indication of the lower metallicity of the Magellanic Cloud stars from the
wavelength displacement of the absorption component: low-metallicity stars
always have a smaller extent (i.e., terminal velocity) 
than their Galactic counterparts.

\section{Synthetic galaxy spectra}

The 75 spectral groups (15 temperature and 5 luminosity classes) were used
to replace the existing solar-metallicity groups of type O in Starburst99.
All other spectral groups are still at the original (near)-solar 
metallicity\footnote{
Note that the existing ``solar'' metallicity library is actually somewhat
subsolar since the sun has a metal overabundance of $\sim$0.2~dex
(Smartt \& Rolleston 1997) with respect to nearby H~II regions where the
 library stars are located. For simplicity we
refer to the existing library as ``solar-metallicity''.}. This
can affect the computed synthetic spectra of an evolving population, 
depending on the IMF and/or the age. The validity range of the low-metallicity
library can be easily assessed: since the lifetime of the least massive 
O stars is $\sim$10~Myr (Schaller et al. 1992), the UV spectrum of a 
{\em single}
stellar population with a standard Salpeter IMF up to 100~\Ms\ will be
determined by the LMC/SMC stars until about 10~Myr. Later on, the original
Galactic B stars will produce the spectrum. 

If stars form {\em continuously},
a near-equilibrium of the UV light is reached after about 10~Myr. In this case,
the synthetic UV spectrum is a mix of the new LMC/SMC stars and
the original Galactic stars, and the consequences of omitting low-metallicity
B stars are not immediately obvious. In general, 
the stellar-wind lines, such as \nv, \siiv,
or \civ, are from massive ($> 40$~\Ms) stars. Therefore the profiles of these
lines predicted by a synthetic model will still be correct for all the ages
considered if B supergiants (the descendants of massive main-sequence stars)
can be neglected. On the other hand, the continuum and many photospheric lines
are due to less massive B stars, for which we use the 
solar-metallicity library. These lines would be too strong in our models
if compared to observations of metal-poor galaxies for which a continuous
model with age older than 10~Myr is deemed appropriate. We performed the
following test: 
The rectified solar-metallicity B stars in the library with temperatures
below 25,000~K were
replaced by a line-free continuum. This simulates the extreme assumption of
zero-metallicity B stars. The O- and early-B (until B0.5) spectra 
were left unchanged, i.e. with
the LMC/SMC stars. In Fig.~\ref{zero} we compare two spectra computed with
this artificial library and with the new LMC/SMC O-star and Galactic B-star
library. Both spectra are for continuous star formation at age 100~Myr and
with a Salpeter IMF. Despite the extreme assumption for the B star
spectra, the differences  are not dramatic. The wind profiles have somewhat
weaker absorptions. This results from adding a featureless continuum in 
this test, whereas in reality even an LMC/SMC B star would have some 
absorption, producing stronger wind lines in the synthetic spectrum. 
The outcome of this test depends quite sensitively on the cut-off temperature
for the replaced B stars. Choosing 30,000 K instead of 25,000~K would
affect the lines of, e.g., Si~III $\lambda$1417 and C~III $\lambda$1427
whose strength is heavily weighted towards the earliest B stars with 
temperatures between 25,000~K and 30,000~K.
We conclude that omission of LMC/SMC stars later than B1
does not affect the interpretation
of the star-formation histories presented in this paper.

The absolute flux
levels of the continuum are always treated correctly since evolutionary
tracks and model atmospheres with the appropriate chemical composition are
used. Therefore the above limitations do not apply.

Details of the input physics and computation technique in Starburst99 are
given in Leitherer et al. (1999). We computed a set of standard model
spectra for evolving starbursts with and without ongoing star formation.
The former and the latter will be referred to as the ``instantaneous'' and 
the ``continuous'' case, respectively. 
A standard Salpeter IMF with slope $\alpha = 2.35$
between 1 and 100~\Ms\ and stellar evolution models with a metallicity
of $\frac{1}{5}$~\Zs\ are used.

A time series for an instantaneous starburst between  0 and 10~Myr is shown
in Fig.~\ref{inst}.
The spectra discussed in this paper are normalized to the unity level. All
spectra were computed in luminosity units as well to permit estimates of,
e.g., the dust obscuration and the star-formation rates. The
corresponding luminosities are not reproduced here, as they are the same as
in Fig.~53d of Leitherer et al. (1999). Both the normalized and the
absolute luminosities are part of the output of Starburst99 and can be
retrieved from the website given at the end of this paper.
The line profiles have the same
qualitative behavior as in the solar-metallicity models discussed by Leitherer
et al. (1995). They gradually strengthen from a main-sequence dominated
population at 0~--~1~Myr to a population with luminous O supergiants at
3~--~5~Myr. Afterwards the emission components decrease until the line spectrum
is due to a population of early-B stars at 10~Myr. At that time, the spectrum
in Fig.~\ref{inst} starts showing contributions from solar-metallicity stars.
\siiv\ is strongest at 4~--~5~Myr. This is different from the solar-metallicity
case where this line reaches its maximum at 3~--~4~Myr (see Fig.~6 of 
Leitherer et al. 1995). The reason is stellar evolution --- not the
difference in  the library spectra. Stars with lower metallicity have longer
main-sequence lifetimes and are hotter. Therefore the entrance into the
phase when \siiv\ becomes strong is delayed with respect to the 
solar-metallicity case. We verified this interpretation by computing a
spectral sequence for the low-$Z$ library with solar-metallicity tracks. In
this case the peak strength of the Si~IV profile is reached at about 4~Myr,
similar to models with \Zs\ library stars.

How different are the new models from the earlier solar-$Z$ spectra? In 
Fig.~\ref{comp} (top)
 we compare the spectral region between 1200 and 1600~\AA\ at
$\frac{1}{4}$~\Zs\ and \Zs. At $t = 0$~Myr, for a zero-age-main-sequence
population, most photospheric and wind lines  are significantly weaker at
low metallicity
than at \Zs. The wind profile of \civ\ has both weaker emission and 
absorption, an immediate consequence of the lower mass-loss rates at lower
$Z$. \nv\ remains essentially unchanged.
\siiv\ is purely interstellar and the difference between the two
spectra results from the different strengths of the interstellar lines in
the library stars. Almost all weak {\em photospheric}
 lines from O stars scale with metallicity.
The strong S~V $\lambda$1502 is a notable exception: there
is little change with metallicity.
Most weak features can be seen at both metallicities. Since the two sets
of library stars were acquired with different telescopes, they cannot be
due to detector noise but are real stellar lines (Nemry, Surdej, \& 
Herniaz 1991). 

The majority of the weak photospheric lines
are due to transitions in highly ionized iron and nickel (Haser et al.
1998). They have typical line widths of tens of \kms\ resulting from
microturbulent velocities of $\sim$20~\kms\ and stellar rotation velocities
of $v \sin i$ between 0  and 100~\kms. 
Therefore all unblended photospheric lines
are unresolved in our library spectra. The comparison in Fig.~\ref{comp}
demonstrates that all photospheric lines become weaker at lower metallicity.
Most lines are too weak to be used as metallicity indicators individually
but the overall blanketing effect as a function of metallicity can provide
a useful estimate of the chemical composition of a galaxy spectrum. The
strongest photospheric lines showing a metallicity dependence are the
Si~III, C~III, and Fe~V blends around 1425~\AA. In the next section, we 
will utilize these lines to estimate the chemical composition of the
high-redshift galaxy \cB.

The 0~Myr old population is a somewhat academic case since a genuine 
zero-age-main-sequence population will hardly be observed in the UV; 
$t = 4$~Myr is more realistic (Fig.~\ref{comp} middle). At that age, supergiants
dominate, and the stellar-wind profiles of  \nv, \siiv,
 and \civ\ are at maximum
strength. There is a mild metallicity dependence in the sense that the
velocities are lower and the absorptions are weaker at lower $Z$. Overall,
the effect is not very dramatic, as expected from the discussion of the
individual stellar spectra in the Section~3. Since most star-formation
regions are observed when supergiants are present, our models predict that
{\em the stellar-wind profiles become  moderately weaker towards lower
metallicity.} The metallicity trends of the  photospheric lines are the
same at 0 and 4~Myr, suggesting that age effects are less important.

The bottom panel of Fig.~\ref{comp} shows a comparison between 
$\frac{1}{4}$~\Zs\ and \Zs\ models for continuous star formation at an
age of 100~Myr. The most striking differences between the two spectra are
the decreased photospheric line-blanketing at lower metallicity and the
weaker absorption component of the \civ\ wind line. The N~V and Si~IV
wind absorption are rather similar, and the emission components of N~V,
and Si~IV, and C~IV are essentially identical. The strongest 
photospheric lines from O- and early-B stars are O~V $\lambda$1371,
Fe~V $\lambda\lambda$1360--80,
Si~III $\lambda$1417, C~III $\lambda$1427, and Fe~V $\lambda$1430. Other
relatively strong and unblended photospheric lines in a starburst spectrum are 
Si~II $\lambda$1265 and $\lambda$1533. However, these lines are not
very useful for our purpose of investigating metallicity effects since
they arise from mid- to late-B stars, which are not included in our low-$Z$
library. In contrast, the former lines around 1370~\AA\ and 1425~\AA\
do not suffer from this shortcoming. 

We define line indices ``1370'' and
``1425'' as the equivalent widths  integrated between 1360 and 1380~\AA\
and between 1415 and 1435~\AA, respectively. These two indices measure
the metallicity-sensitive spectral blends from O- and early-B stars 
discussed above. They are plotted in Fig.~\ref{index} for constant 
star-formation models with Salpeter IMF between 1 and 100~\Ms. Their weak
age dependence and relatively strong sensitivity to the line blanketing
makes them useful indicators of the metallicity. The metallicity
dependence in Fig.~\ref{index} is caused by two effects: the spectral
features are weaker in metal-poor individual stars {\em and} stellar
evolution models are metallicity dependent as well. In addition, 
systematic errors in the models are not entirely negligible. The
continuum normalization and the definition of the library-star grid
may have metallicity dependent biases. For these reasons, the prudent
approach would be to compare the actual spectrum
over a  larger wavelength range to the observations, rather than limiting
the available information by measuring equivalent widths only. 
We estimate the
{\em systematic} uncertainty of each point in Fig.~\ref{index} to be
about $\pm 0.1$~\AA, with a possible time dependence. 
Therefore any trend with time is not significant. 

In addition,  there will be a possibly even larger error 
due to the adopted continuum location. This
error is hard to quantify. Therefore it is important to understand how
the continuum is defined. Evolutionary synthesis is driven by stellar
evolution models and their predicted dependence of stellar effective
temperatures and bolometric luminosities with time. The emergent
fluxes at any wavelength point come from model atmospheres. Therefore
our definition of the continuum is identical to that used in the
model atmospheres. Since theoretical atmospheres know a priori the
amount of line-blanketing at each wavelength point, the continuum is
not derived by just fitting an unbiased power law to the fluxes. Rather,
we first select wavelength intervals which are line-free according to the
atmosphere models. The wavelength points for
the fits are 1150~\AA, 1280~\AA, 1310~\AA, 1360~\AA, 1430~\AA, 1480~\AA,
1510~\AA, 1580~\AA, 1630~\AA, 1680~\AA, 1740~\AA, 1785~\AA, 1820~\AA, and
1875~\AA. This is done for every single position in the 
Hertzsprung-Russell diagram.
These wavelength regions are assumed to be pure continuum and to be 
time-independent. This assumption is true for most of the time. If weak
photospheric lines affect the chosen continuum points at certain time steps,
the predicted continuum of the synthetic spectrum may be slightly offset
from the unity level. Then we fit a smooth curve through these points
and replace the original model atmosphere by the smooth curve. Equivalently,
one can describe this as removing any spectral lines and generating 
a ``pseudo-atmosphere'' at an optical depth
where free-bound, free-free, and e$^-$-scattering processes dominate. The second
step is to link the normalized, {\em line-blanketed} library spectra to the
luminosity calibrated {\em line-free} model atmospheres. The library spectra
were rectified using the same prescription, i.e. we fitted a curve only to
those spectral regions which we knew a priori are line-free. This method
is different from, e.g., the prescription in Walborn et al. (1985) who
rectified the continuum to unity based on the empirical
absence of features in a spectral region, rather than the theoretical 
prediction the line-blanketing.
 To summarize, the continuum level in the model spectra
depends (i) on
the fitting points of the ``pseudo-atmosphere'' and (ii)
 on the adopted normalization of the library spectra. 
The two-step approach is
required due to the reddening of the model spectra. In the absence of 
reddening, we could have simply linked the luminosity calibrated library
spectra with stellar evolution models. It is crucial to recall these steps
when comparing weak lines between models and observations. 
The continuum definition is a major source of uncertainty both in the
observations and in the synthetic models.

Spectra of low-metallicity populations with continuous star formation from 0
to 500~Myr are shown
in Fig.~\ref{cont}. The properties of this sequence hold no
surprises, given what was discussed before. Spectra with continuous
star formation are relevant for the modeling of the integrated light of
galaxies, such as, e.g., Lyman-break galaxies at high redshift. For
star-formation timescales larger than about 50~Myr, only the \civ\ stellar-wind
line can be readily recognized as stellar; all other stellar lines are 
 blended with interstellar absorption lines. Apart from \civ, the
next strongest stellar features are \nv, \siiv, and  the Si~II and Si~III 
multiplets around 1300~\AA\ (see de~Mello et al. 2000). Sufficiently
high spectral resolution is required observationally in order to 
 separate  these features from interstellar Lyman-$\alpha$,
\siiv, and O~I~+~Si~II $\lambda$1303. Other stellar photospheric lines
seen in these spectra are Si~III $\lambda$1417, C~III $\lambda$1427, and
S~V $\lambda$1502.

\section{Comparison with the UV spectra of NGC~5253 and MS~1512-cB58}

The first application of the new low-metallicity library to UV observations
of a star-forming galaxy was done by Tremonti et al. (2000) in their study
of the metal-poor galaxy NGC~5253. This galaxy hosts numerous young super star 
clusters with an oxygen abundance of about $\frac{1}{6}$~\Zs\ 
(Calzetti et al. 1997). Since the analysis of the UV spectra was performed
by Tremonti et al., we will not repeat their study but rather use their
results to highlight the properties of the low-metallicity synthesis models.

In Fig.~\ref{n5253} we compare the average spectrum of 8 clusters in
NGC~5253 to synthetic models at solar and $\frac{1}{4}$~solar metallicity.
The two models have the same parameters, except for the metallicity. The
metallicity adopted in the evolutionary tracks is \Zs\ and 
$\frac{1}{5}$~\Zs\ for the solar and the LMC/SMC library, respectively. A
standard Salpeter IMF between 1 and 100~\Ms\ was used. The starburst is
extended, with stars forming continuously for 6~Myr. This star-formation
history is appropriate for a superposition of 8 clusters with ages between
1 and 7~Myr. However, the precise age does not strongly affect the computed 
spectrum for the case of continuous star formation (see Fig.~4 of Leitherer
et al. 1995). The agreement between the average cluster spectrum and the
modeled low-$Z$ spectrum is excellent. The stellar-wind features \nv,
\siiv, and \civ\ are reproduced extremely well, except for the narrow 
interstellar contributions. We do not expect to match the interstellar
lines in the observed spectrum for two reasons. First, the library stars have
a mean reddening of $E(B-V)=0.11$, whereas the cluster reddening is 
$\sim$0.25. Consequently, the column density of the interstellar lines
in NGC~5253 is higher than in the LMC/SMC library stars, and the line strengths
of not too deeply saturated lines should be larger in NGC~5253. Second, and
more important, Heckman \& Leitherer (1997) suggested that the interstellar
lines in NGC~1705 are broadened due to macroturbulence and blending of
multiple components due the energy input
from stellar winds and supernovae in starbursts. Generally, the interstellar
medium in starburst galaxies is more turbulent than in the Milky Way and
the Magellanic Clouds so that even  saturated lines are stronger in
starbursts.

The corresponding stellar-wind lines produced by the \Zs\ model spectrum are
a worse fit to the observations. Particularly discrepant are the
blue absorption wings in \siiv\ and \civ, which are much too strong in
the models. Reducing the number of the most massive stars with a steeper
IMF would somewhat alleviate the problem, but at the price of a worse fit
to the emission components. The ratio of the absorption to the emission 
strengths is therefore a sensitive indicator of the metallicity. We expect
this relation to be tighter for \civ\ than for \nv\ and \siiv\ due to the
previously discussed ionization effect. As opposed to C$^{3+}$, N$^{4+}$
and Si$^{3+}$ are trace ions in hot-star winds, and changes in the ionization
balance with metallicity complicate the direct metallicity dependence. 
  
Photospheric line-blanketing can be seen over the entire spectral range plotted
in Fig.~\ref{n5253}. The most prominent examples are C~III $\lambda$1428 and 
S~V $\lambda$1502. Most photospheric lines are weak in comparison with
wind lines so that metallicity effects are strong. Obviously the 
low-metallicity spectrum provides a much better fit to the observations
than the \Zs\ model. The agreement between the modeled and observed
photospheric lines suggests that the metallicity used in the model is
appropriate for NGC~5253. Vice versa, the photospheric lines (and of course
the \civ\ wind profile) can be used {\em to determine the metallicity} if
the other stellar properties are known independently. 

As a second test, we apply our low-metallicity library to the restframe-UV
spectrum of the $z=2.72$ galaxy \cB. This object was originally discovered
in the CNOC cluster redshift survey by Yee et al. (1996) and subsequently
found to be a lensed star-forming galaxy (Seitz et al. 1998). Its 
star-formation history and general properties were explored by 
Ellingson et al. (1996), de~Mello et al. (2000), Pettini et al.
(2000), and Teplitz et al. (2000). Both de~Mello et al. and Pettini et al.
modeled the restframe-UV spectrum of \cB\ with the original \Zs\ library
in Starburst99. The new low-$Z$ library allows us to model the UV spectrum 
in a truly self-consistent way.

The chemical composition of \cB\ is known only approximately. 
Due to its redshift, a traditional emission-line analysis is not feasible
since some of the key diagnostic lines, like [O~III] $\lambda4363$,
 are inaccessible from the ground and/or
are simply too faint to be observed. Nevertheless, Teplitz et al. (2000)
were able to estimate an oxygen abundance $\frac{1}{3}$~\Zs\ from near-IR
spectroscopy, using the approximate R$_{\rm{23}}$ method (e.g., Kobulnicky,
Kennicutt, \& Pizagno 1999). This value agrees with other independent 
determinations: 
Leitherer (1999) used the relation between the  equivalent widths of strong
UV lines and the
oxygen abundance to suggest LMC/SMC-like abundance in \cB. Pettini et al.
(2000) analyzed  weak interstellar
absorption lines of Si~II, S~II, and Ni~II to infer an abundance of
about $\frac{1}{4}$~\Zs. 
Heap et al. (1999) modeled the stellar-wind lines in the spectrum of \cB\
and found a metallicity like that in the SMC.
Here we propose a new technique: 
{\em stellar photospheric lines} in the UV spectrum. This method was 
successfully used
to estimate the metal abundance in hot stars by Haser et al. (1998) but has
not been applied before to integrated spectra of star-forming galaxies.
Photospheric lines in the UV may not allow abundance determinations as
precise as those from optical nebular emission lines but they can certainly be
very useful when the optical wavelength range is inaccessible to observations. 

The key spectral region between 1400 and 1440~\AA\ is shown in 
Fig.~\ref{abundances}. The ``1425'' index defined in the previous section
is in this wavelength region.
The spectrum of \cB\ is compared to two model spectra
at \Zs\ and $\frac{1}{4}$~\Zs, both calculated for continuous star formation
with age 100~Myr and Salpeter IMF from 1 to 100~\Ms. This corresponds to 
the parameters derived by Pettini et al. (2000). However, the strength of
the photospheric lines --- as opposed to the wind lines --- does not 
strongly depend on these specific parameters, as demonstrated in 
Fig.~\ref{index}. The ``1425'' index at $\frac{1}{4}$~\Zs\ and \Zs\
metallicity has a value of 0.4 and 1.5~\AA, respectively. It is obvious that the
model with solar composition is a rather poor match to the observed spectrum.
The low-metallicity model fits the observations rather well. The comparison
in Fig.~\ref{abundances} suggests that the newly formed stars in \cB\ 
have a chemical composition similar to that of the LMC/SMC library stars.

A full comparison between the models and observations of \cB, including
the stellar-wind lines is performed in Fig.~\ref{cB58}. The structure of
this figure is identical to that for NGC~5253 (Fig.~\ref{n5253}). The
models shown are those of Fig.~\ref{abundances}. {\em Before comparing models
and observations, one should recall that the narrow interstellar components,
including those superposed upon the broad stellar-wind lines, are not 
intended to be reproduced by the models.} The overall spectral fit
with the low-$Z$ library is much improved over that with solar composition.
Apart from the better fit to the photospheric lines discussed before, there
is a significant improvement in the modeling of the wind absorption
components. At solar chemical composition, a blue absorption wing to 
\siiv\ is predicted but not observed. This wing is not present at 
$\frac{1}{4}$~\Zs. Even more strikingly, the gross mismatch between the
observed and theoretical \civ\ absorption at \Zs\ almost disappears at
low metallicity --- a clear indication for the lower than solar metallicity
of \cB. The relative strength of the \civ\ absorption to the emission of
this and other wind lines is useful as a metallicity indicator: the absorption
decreases with lower $Z$, all other galaxy properties being equal.

The properties of the stellar content of \cB\ derived with the low-metallicity
library are not significantly different from that derived before with the
\Zs\ library. Assuming constant star formation, the stellar-wind profiles
decrease in strength very slowly and essentially monotonically after about
5~Myr (see Fig.~\ref{cont}). In this case, only the IMF determines the
relative proportion of O and B stars, and consequently the line strength. A
flatter IMF produces stronger emission and absorption. The \civ\ profile
is a very sensitive diagnostic: the model somewhat underpredicts the emission
and overpredicts the wind absorption blueward of the deep interstellar
line. Changing the IMF would always improve the fit to either component at
the expense of the other. A Salpeter IMF is a reasonable trade-off in order
to achieve a reasonable overall fit.

We experimented with more complicated star-formation histories in order to
improve the fit. The C~IV profile in \cB\ has relatively strong emission
relative to the absorption. The sequence in Fig.~\ref{cont} indicates that
no continuous model at any age would fully match the observations. A flatter
IMF would produce a {\em ratio} required by the observations but emission
and absorption would be much too strong. Inspection of Fig.~\ref{inst}
suggests a possible scenario: an instantaneous starburst produces luminous
O supergiants at 3~--~5~Myr after the burst. Such stars have the maximum
emission/absorption ratio possible. If such a burst is combined with an
older population in order to dilute both the emission and absorption, an
excellent fit to the observations can be constructed. We verified that this
is indeed the case: a 4~Myr old burst with a flat IMF, superposed upon
a 1~Gyr old population with constant star-formation and a steep IMF provide
a much better fit than the model shown in Fig.~\ref{cB58}. However, this is
a somewhat academic exercise and would not add new insight already gained
from the standard model: \cB\ has a metallicity of $\frac{1}{4}$~\Zs, and is
currently experiencing a powerful starburst producing massive stars down
to at least 10~\Ms\ and following a more or less standard IMF.

Could  the mismatch between the observed and theoretical \civ\ profiles
be caused by an abundance effect? Decreasing the carbon
abundance will affect mostly the absorption and leave the emission relatively
unchanged. This is the trend we found from Galactic to LMC/SMC abundances,
and we suspect --- but have no observational proof --- that the trend
continues. Our test using zero-metallicity B stars provides tentative
support for this hypothesis (see Fig.~\ref{zero}). The C~IV absorption is 
indeed weaker in the test spectrum of Fig.~\ref{zero}. On the other hand,
zero-metallicity B stars generate no excited Si~II $\lambda$1265, a line
which is clearly present in \cB, and which is nicely reproduced by our 
model in Fig.~\ref{cB58}.
The carbon abundance of the library stars is the result of the
chemical history of the Magellanic Clouds. A priori there is no reason 
why the chemical evolution of \cB\ should be identical to that of the
Magellanic Clouds. Since \cB\ is likely to be at a relatively early 
stage of evolution, given its redshift, fewer generations of low-
and intermediate-mass stars could have formed. Since these stars are
responsible for the carbon production, chemical evolution models predict
an underabundance of carbon relative to elements predominantly
formed in massive stars, like oxygen (Bradamante, Matteucci, \& D`Ercole
1998). However, the C~III $\lambda$1427 line in \cB\ shows no peculiarity
(see Fig.~\ref{abundances}). Its strength, both in absolute terms
and relative to Si~III $\lambda$1417
and Fe~V $\lambda$1430, is approximately the same as in the model spectrum.
This suggests that intermediate-mass stars may have had sufficient time
to evolve and release carbon in \cB.

\section{Conclusions}

The new low-metallicity library complements our previous model spectra 
based on solar-metallicity stars and allows an important extension of
the parameter space available for study. Galaxies with metallicity close
to that of the SMC, and with correspondingly low mass and luminosity,
can be modeled. We find that metallicity affects the UV spectrum, but
not in a dramatic way. Therefore previous studies, which compared  
observed spectra of metal-poor
galaxies to models with solar-metallicity library stars, are still 
approximately valid.

What improvements could be made in the future? Pushing the library towards
significantly lower metallicities is out of reach. The only other very
metal-poor galaxy
in the Local Group with a significant massive-star population is IC~1613
(Armandroff \& Massey 1985). The metallicity of this irregular galaxy
is half that of the SMC (Cole et al. 1999). At a distance of 400~kpc the
brightest O stars are 3 magnitudes fainter than their cousins in the SMC.
The corresponding increase of the observational effort to define an
optical sample and subsequently accumulate
a body of UV spectra can hardly be justified in view of the incremental
increase of the parameter space. 

Two significant enhancements of the library can be accomplished at a modest
observational expense. First, extension of the spectral-type coverage to
late-B stars would allow us to apply the spectral synthesis analysis to
galaxies at evolutionary stages 
when stars with masses as low as 5~\Ms\ become important. Spectral features
of such stars are already contributing to the spectrum of, e.g., \cB, and
they become dominant in more evolved systems such as post-starburst galaxies.
A second enhancement is the extension of the wavelength coverage to longer
wavelengths, both at low and high metallicity. 
B stars in particular have important diagnostics between 1600
and 3000~\AA, such as Al~III $\lambda$1860 and Si~III $\lambda$1892
(Walborn, Parker, \& Nichols 1995). 

The synthetic UV spectra at low metallicity are part of Starburst99
 (Leitherer et al. 1999). This application allows users to generate tailored
 models using a web based interface at \linebreak
 {\tt www.stsci.edu/science/starburst99/}.
 The original Starburst99 package has been updated to incorporate the models
 described in this paper. Users can generate their models via the ``Run a
 simulation'' page by specifying the low-metallicity library. In addition,
 a set of model spectra with standard stellar-population parameters is
 available for download from the same website.

\acknowledgments
Jo\~ao Le\~ao acknowledges support from the Space Telescope Science Institute
Summer Student Program. We are grateful to Nolan~Walborn for providing comments
on an earlier version of the manuscript.
This work was supported by
HST grant GO-07437.01-96A from the Space Telescope Science Institute,
which is operated by the Association of Universities for Research in
Astronomy, Inc., under NASA contract NAS5-26555.





\clearpage


\figcaption[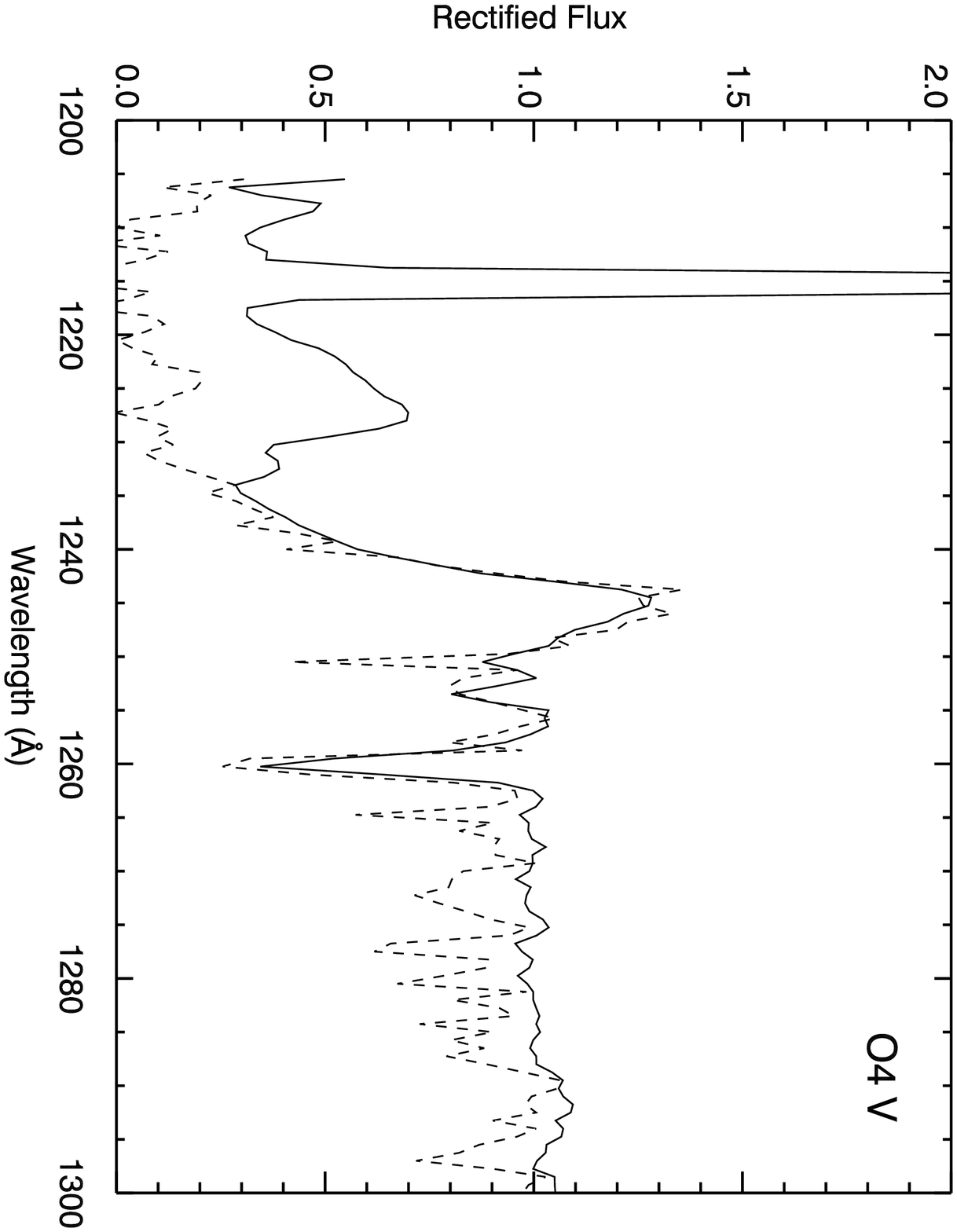]{\label{o4v} Comparison between the low- 
(solid) and solar-metallicity (dashed) spectra within the
O4~V spectral group.
Top: \nv; middle: \siiv; bottom: \civ.}

\figcaption[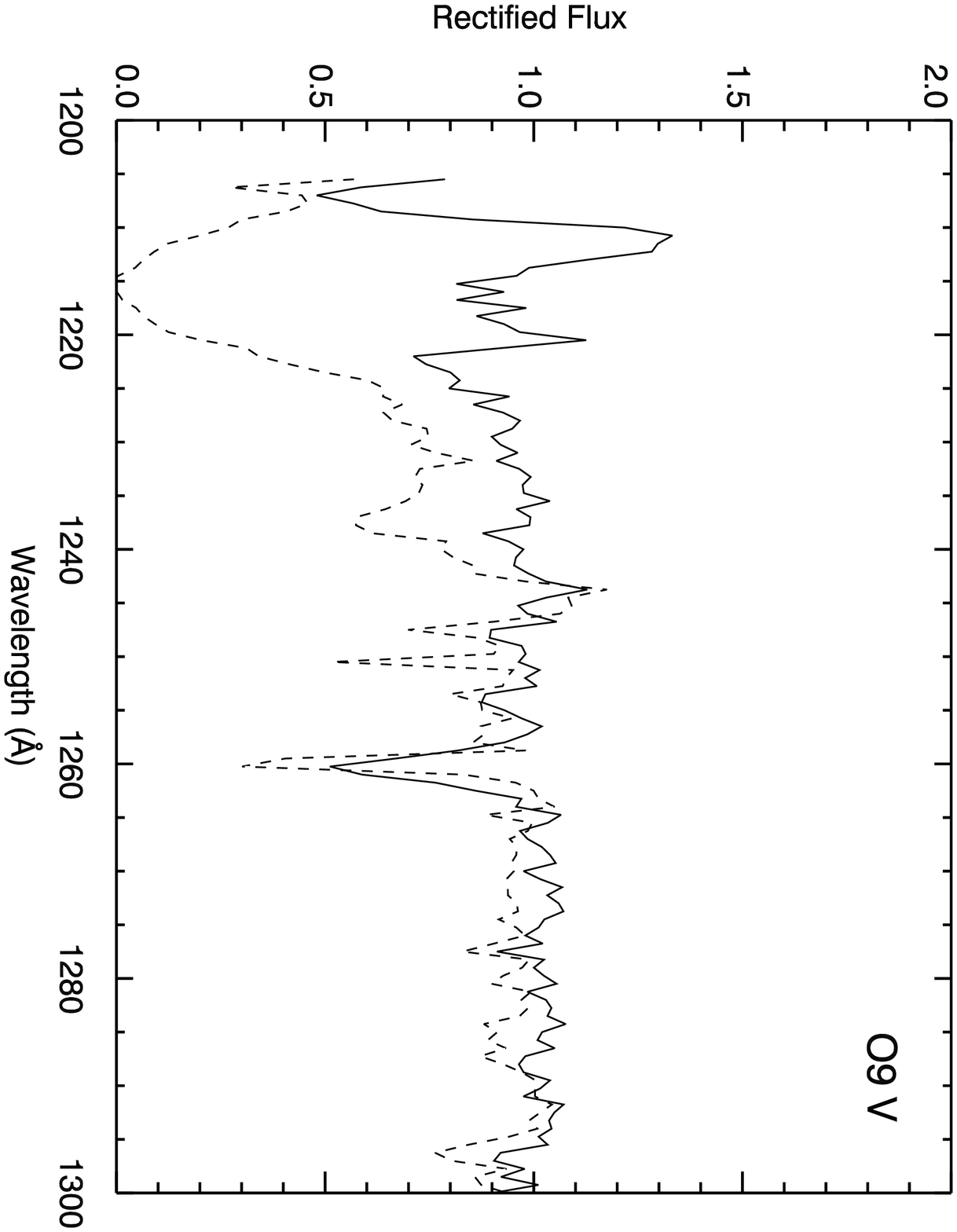]{\label{o9v} Same as Fig.~\ref{o4v} but
for spectral group O9~V.}

\figcaption[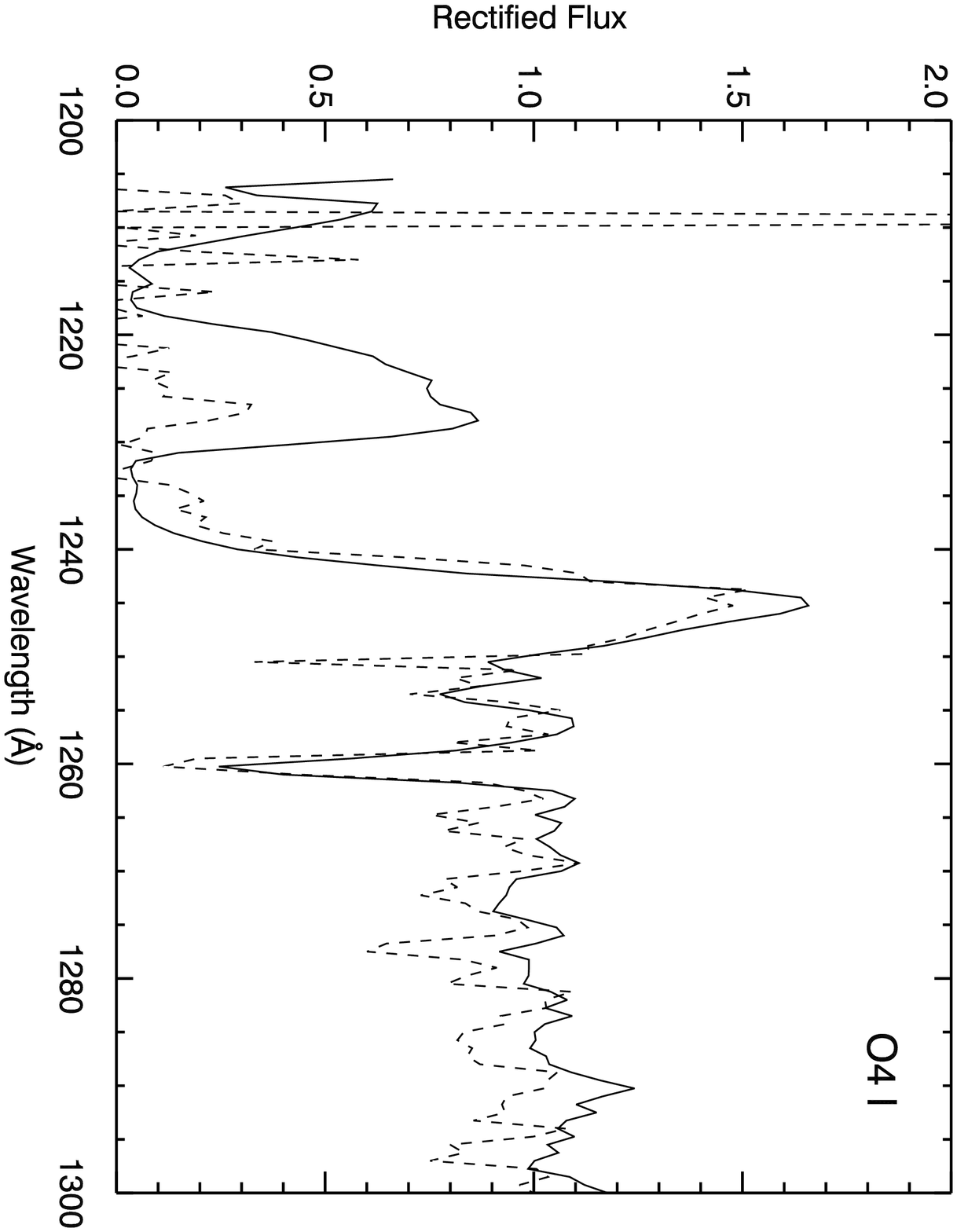]{\label{o4i} Same as Fig.~\ref{o4v} but
for spectral group O4~I.}

\figcaption[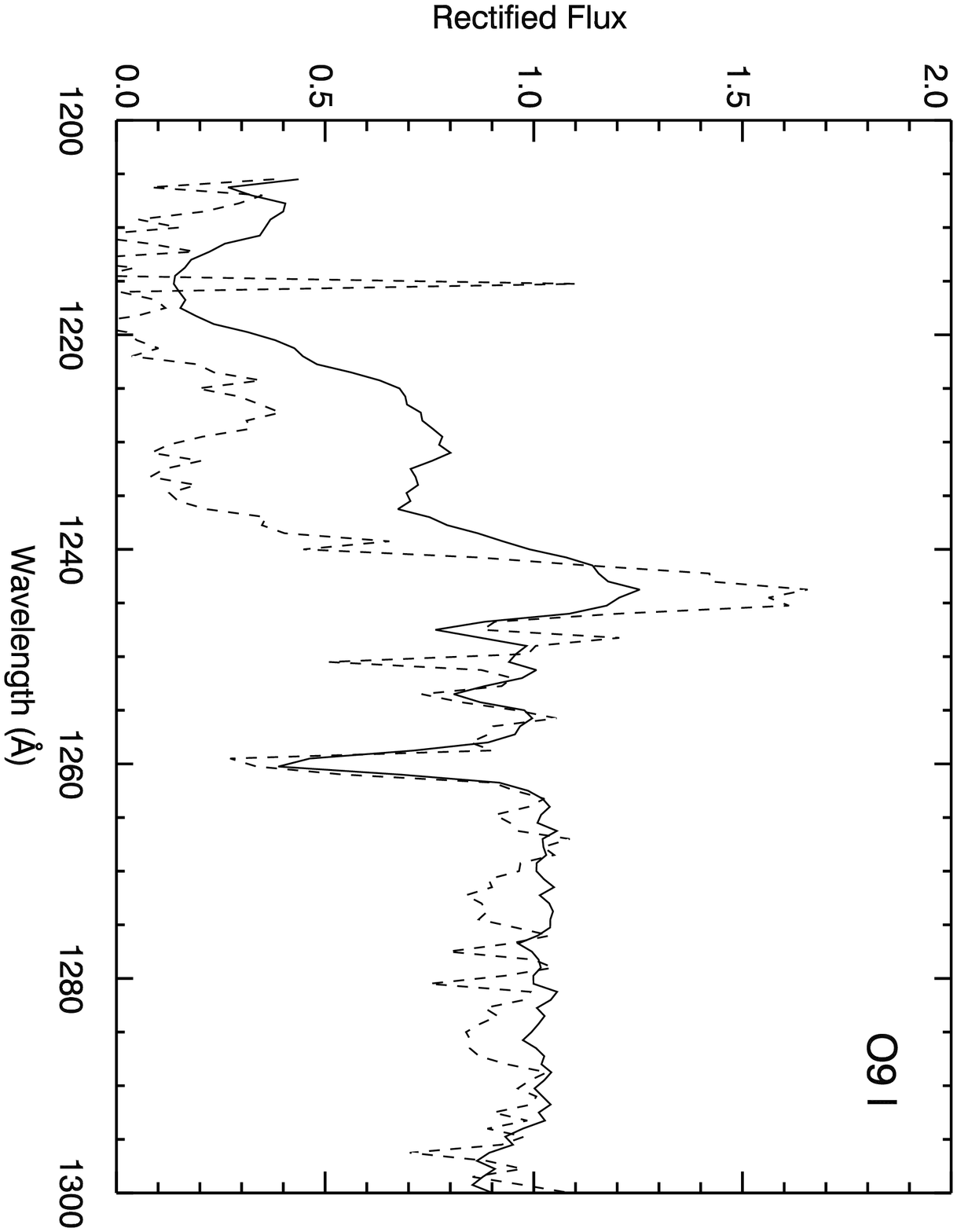]{\label{o9i} Same as Fig.~\ref{o4v} but
for spectral group O9~I.}

\figcaption[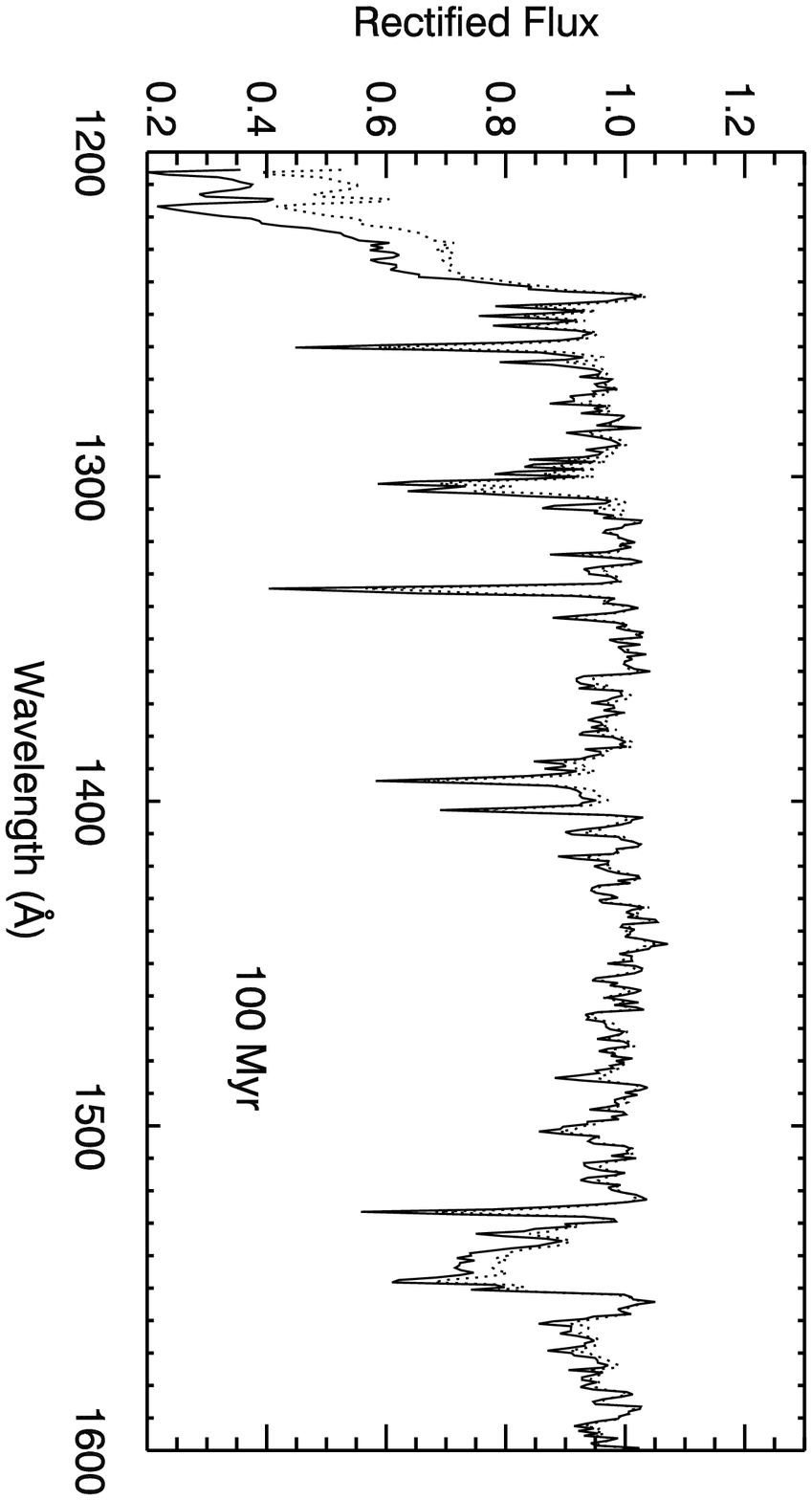]{\label{zero} Comparison between a model
spectrum at $\frac{1}{4}$~\Zs\ computed with the new library (solid
line) and with an artificial library having line-free B star spectra
(dotted). Continuous
star formation, Salpeter IMF with \mlow~=~1~\Ms\ and \mup~=~100~\Ms,
age 100~Myr.}

\figcaption[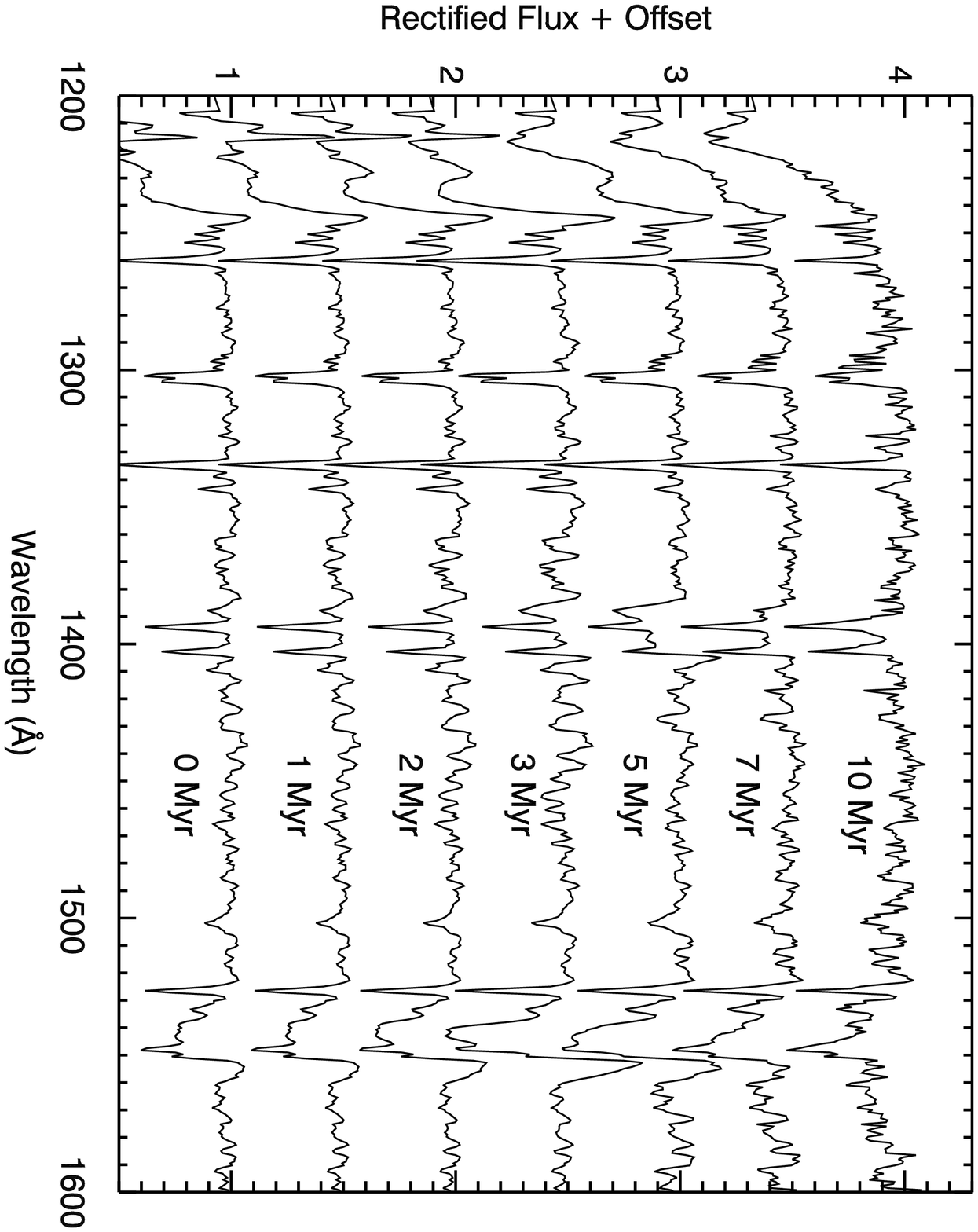]{\label{inst} Evolution of a synthetic 
 UV spectrum between 0 and 10~Myr. Parameters:
instantaneous starburst, $\frac{1}{4}$~\Zs, Salpeter IMF
with \mlow~=~1~\Ms\ and \mup~=~100~\Ms.}

\figcaption[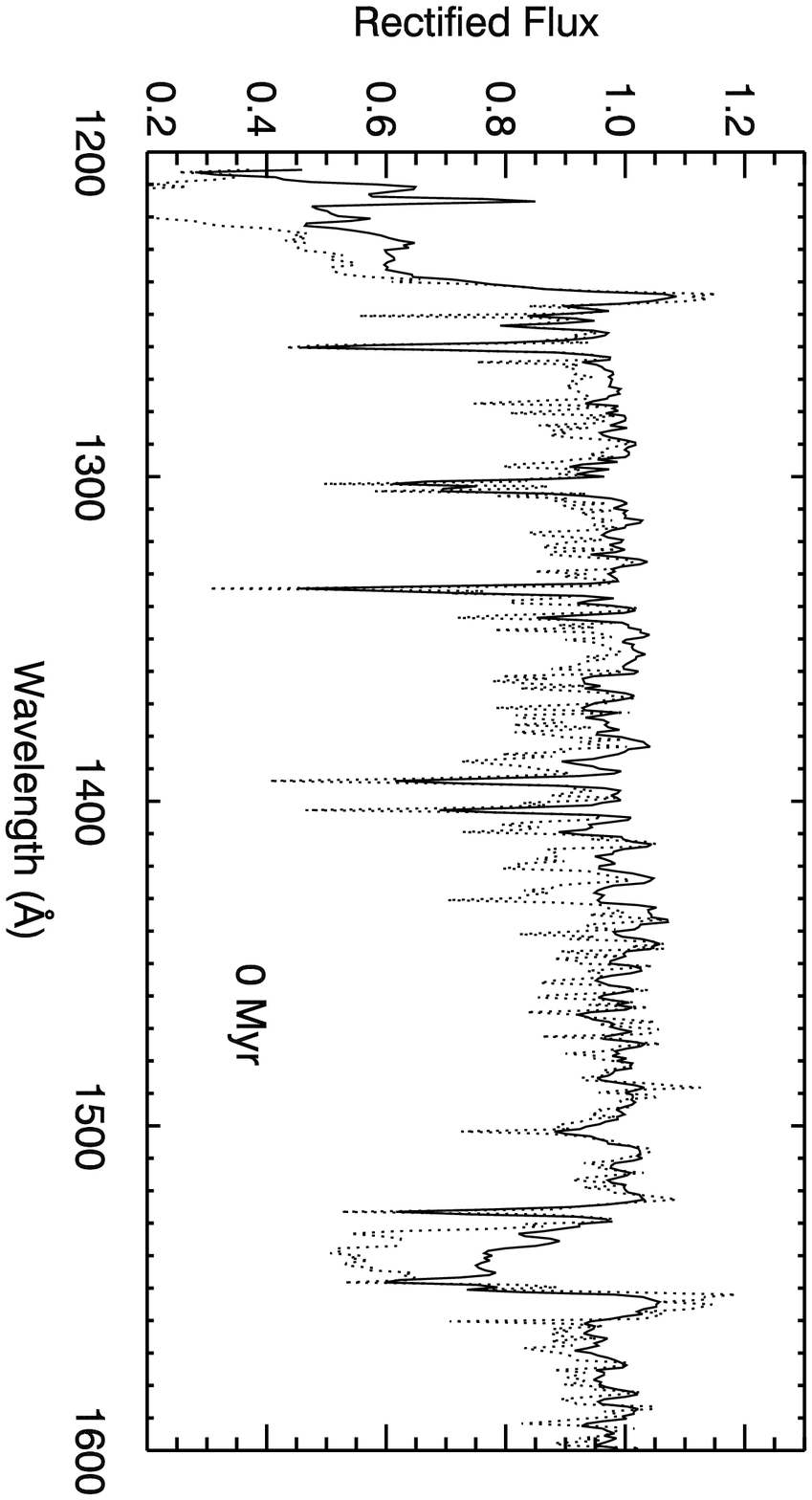]{\label{comp} Comparison of three synthetic
spectra at 0~Myr (top), 4~Myr (middle), and 100~Myr (bottom). Solid: 
$Z = \frac{1}{4}$~\Zs; dotted: $Z$~=~\Zs;
Salpeter IMF between 1 and 100~\Ms; instantaneous starburst for the 0 and
4~Myr models, and continuous star formation for the 100~Myr model.}

\figcaption[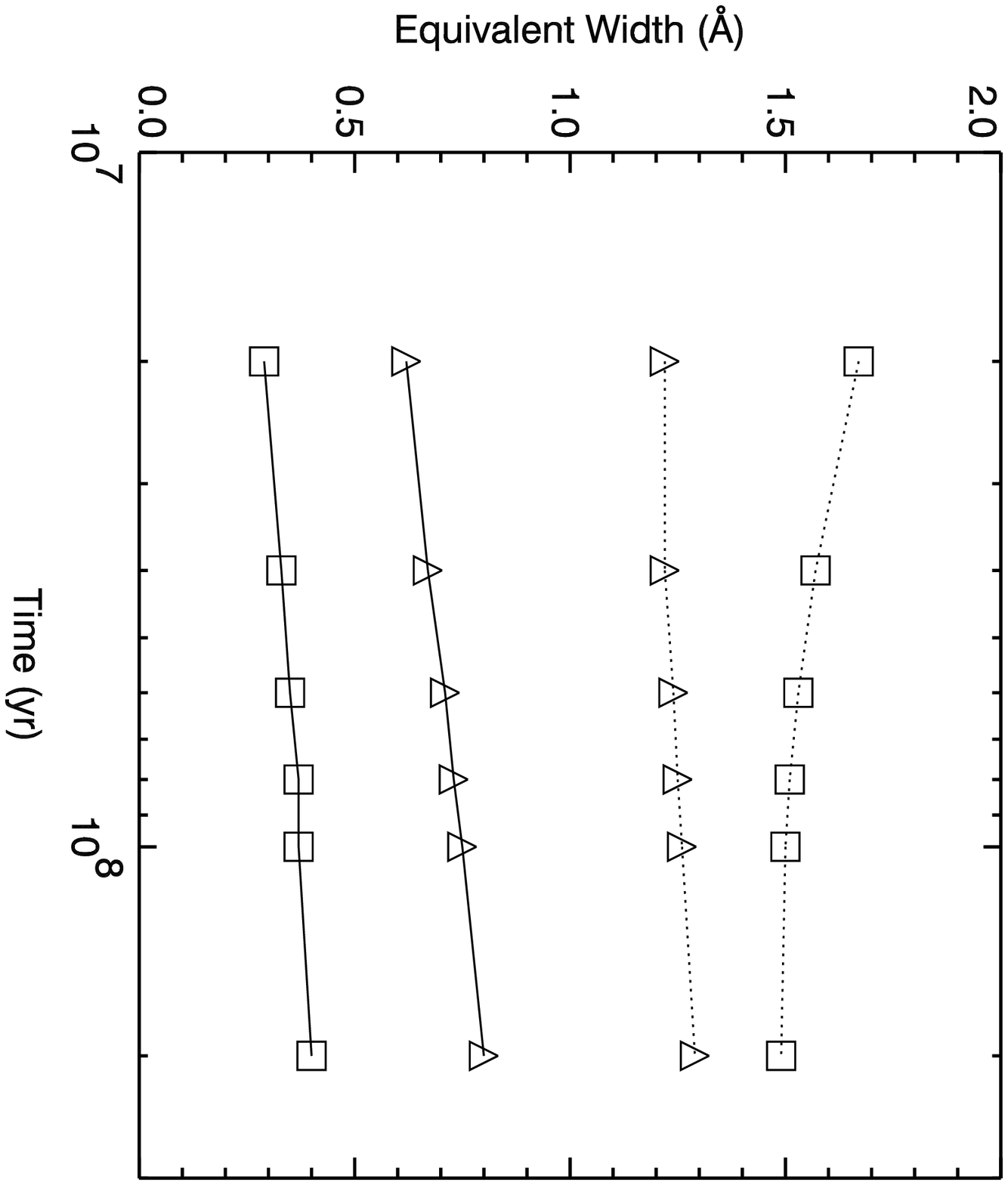]{\label{index} Equivalent width of the ``1425'' index
(squares) and the ``1370'' index (triangles) vs. time. The symbols are
connected by solid and dotted lines for $\frac{1}{4}$~\Zs\ and \Zs\
metallicities, respectively. Continuous star formation with Salpeter IMF
between 1 and 100~\Ms.}

\figcaption[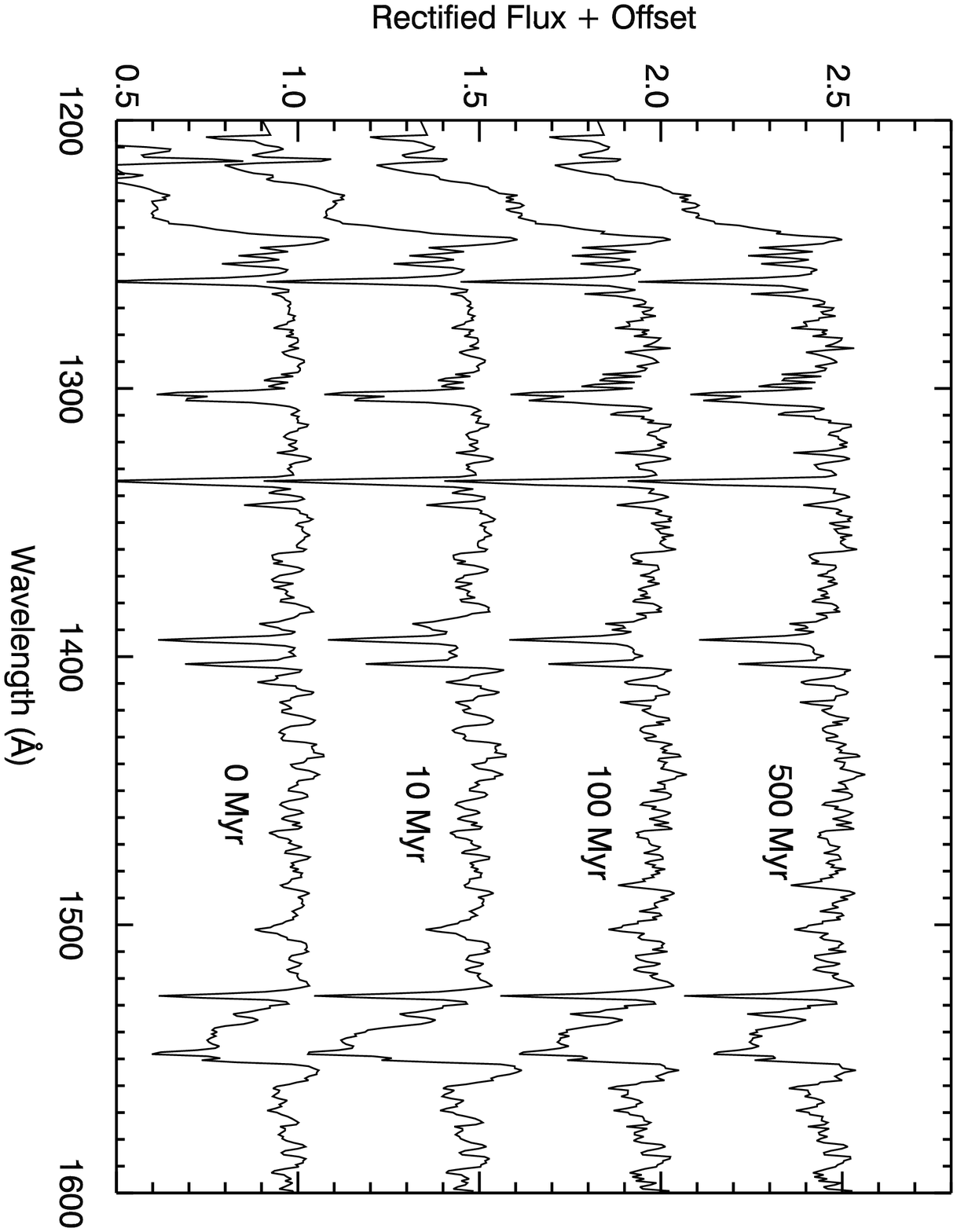]{\label{cont} Evolution of a synthetic 
 UV spectrum between 0 and 500~Myr. Parameters:
continuous star formation, $\frac{1}{4}$~\Zs, Salpeter IMF
with \mlow~=~1~\Ms\ and \mup~=~100~\Ms.}

\figcaption[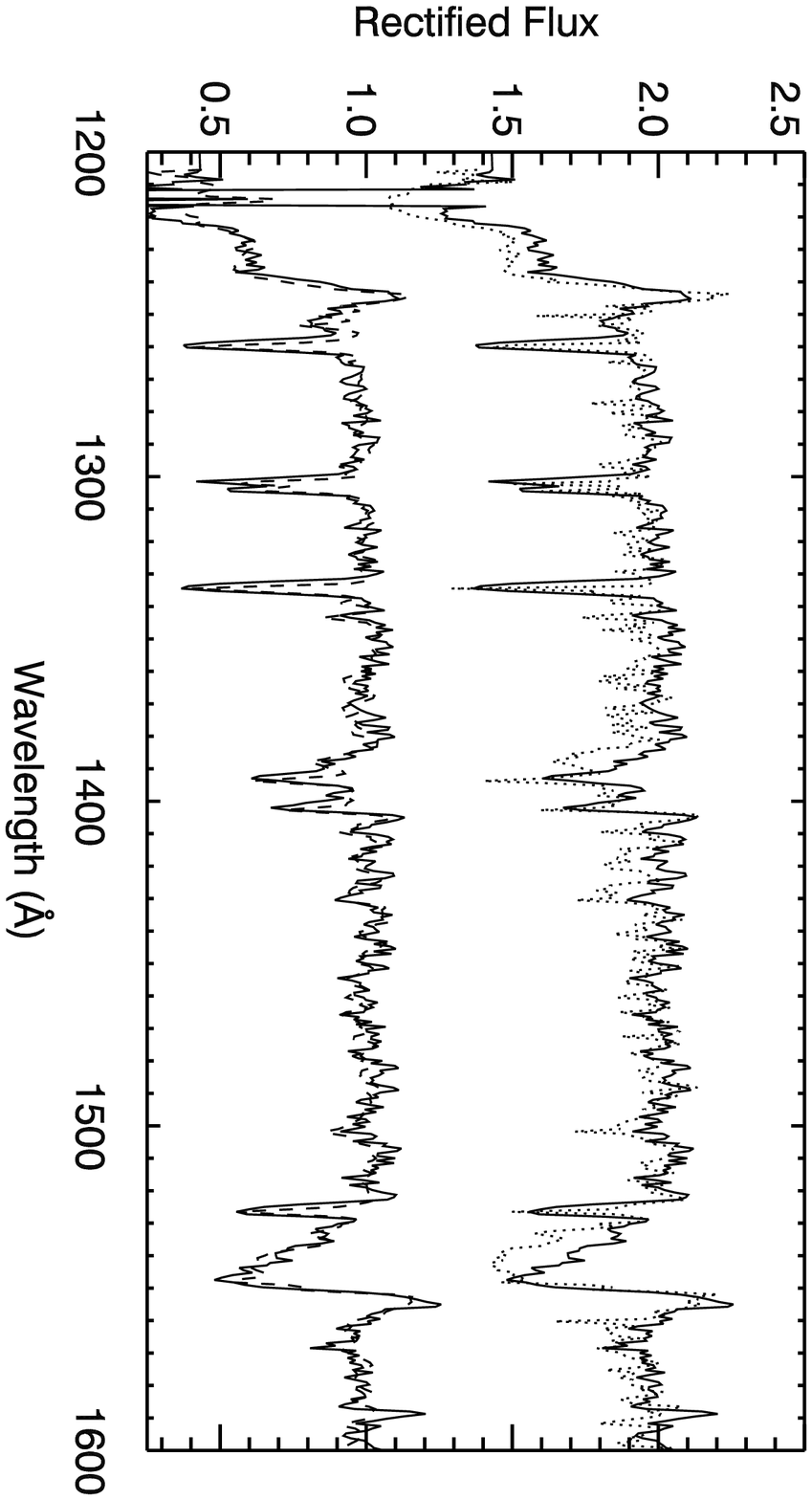]{\label{n5253} Comparison between the average
spectrum of 8 starburst clusters in NGC~5253 (solid lines) and two
synthetic models at $Z= \frac{1}{4}$~\Zs\ (lower; dashed) and $Z$~=~\Zs\
(upper; dotted). The models have continuous star formation, age 6~Myr,
and Salpeter IMF between 1 and 100~\Ms.}

\figcaption[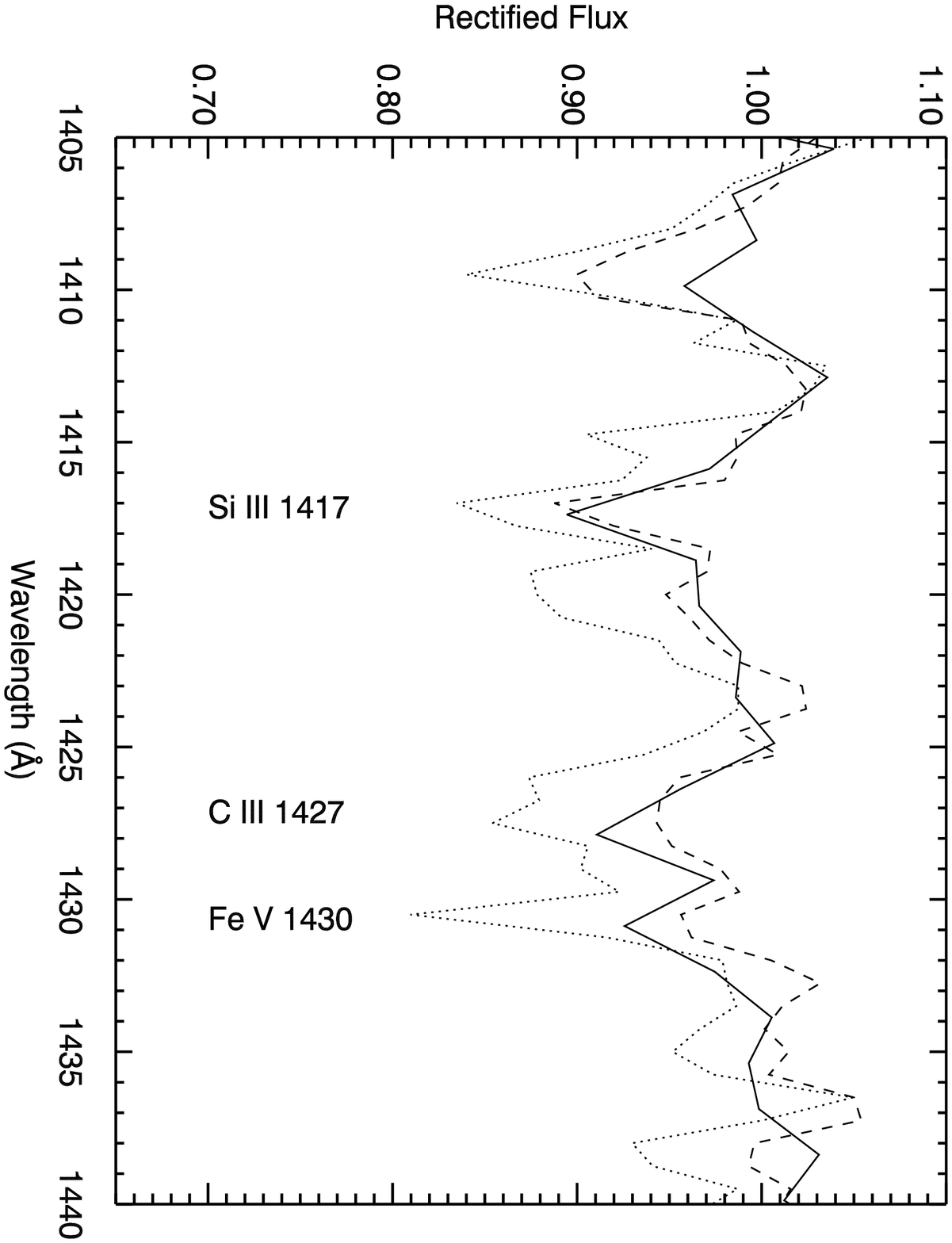]{\label{abundances} Spectral region around
the photospheric features Si~III $\lambda$1417, C~III $\lambda$1427,
and Fe~V $\lambda$1430. Solid line: observed spectrum of MS~1512-cB58;
dashed: model with $Z=\frac{1}{4}$~\Zs; dotted: model with $Z$~=~\Zs
Model parameters: continuous star formation with age 100~Myr, Salpeter
IMF between 1 and 100~\Ms.} 

\figcaption[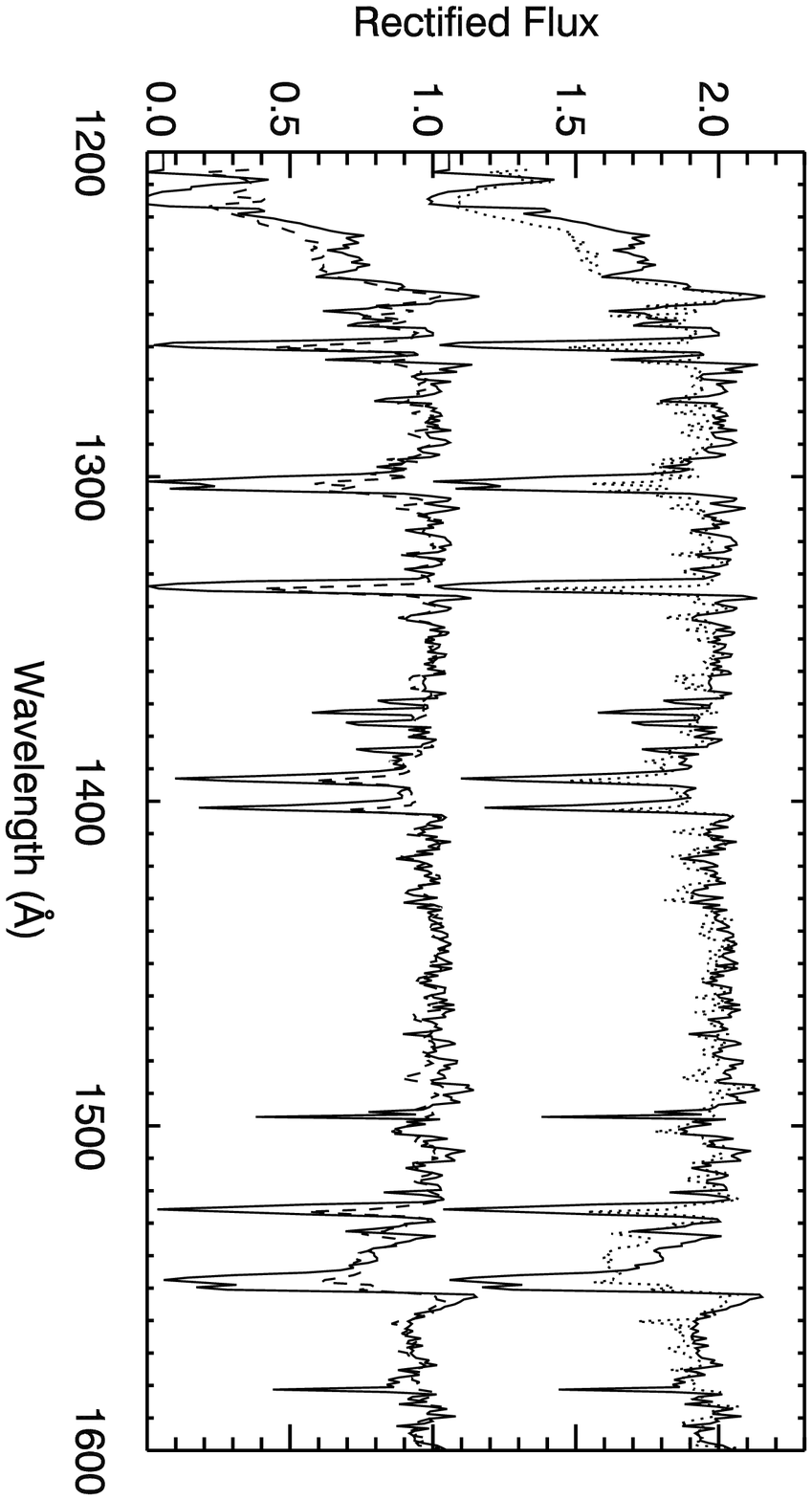]{\label{cB58} Comparison between the observed
spectrum of \cB\  (solid lines) and two
synthetic models at $Z= \frac{1}{4}$~\Zs\ (lower; dashed) and $Z$~=~\Zs\
(upper; dotted). The models have continuous star formation, age 100~Myr,
and Salpeter IMF between 1 and 100~\Ms.}

\clearpage

\psfig{figure=o4v_nv.ps,angle=90,width=9cm,height=6.5cm}
\medskip
\psfig{figure=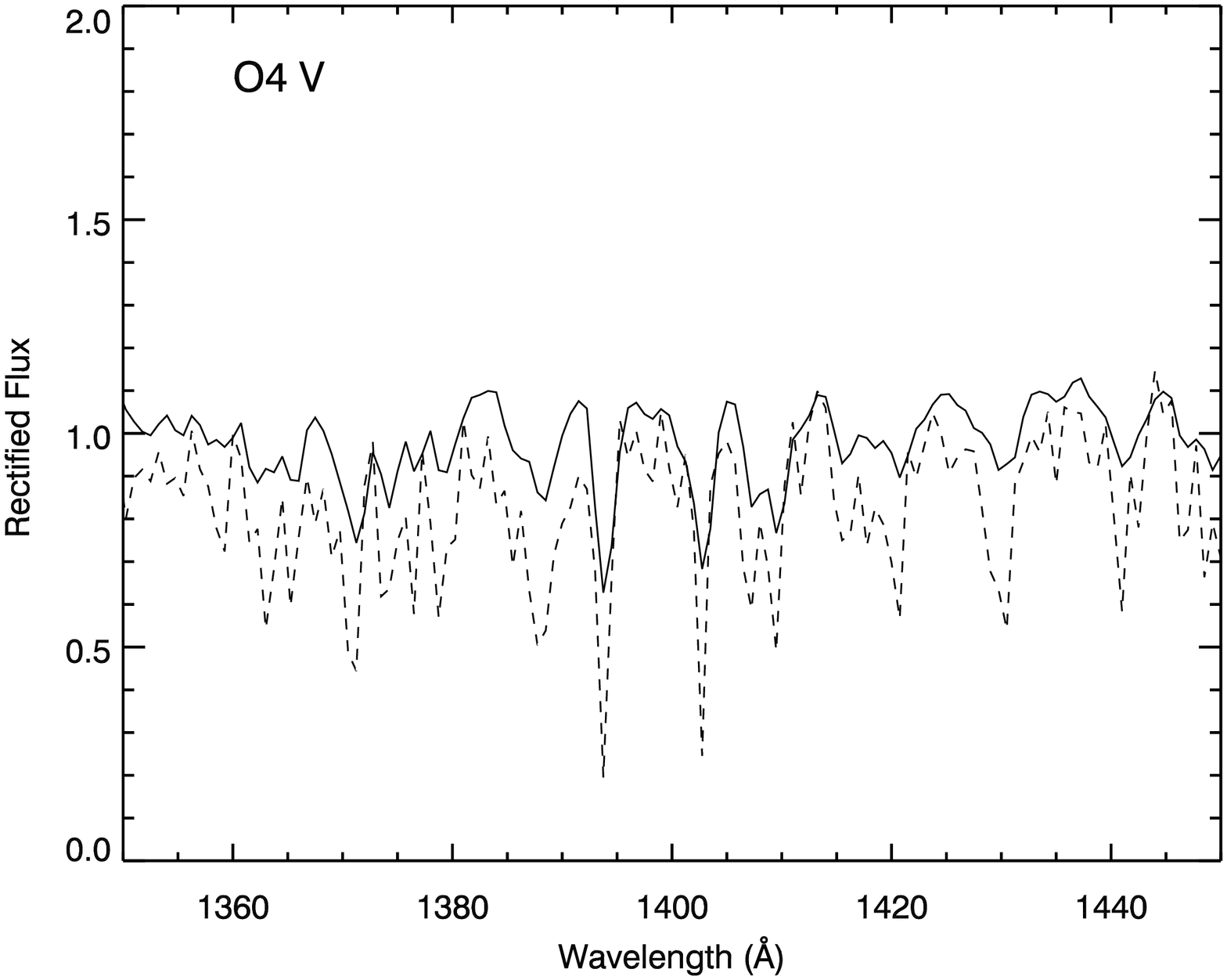,angle=90,width=9cm,height=6.5cm}
\medskip
\psfig{figure=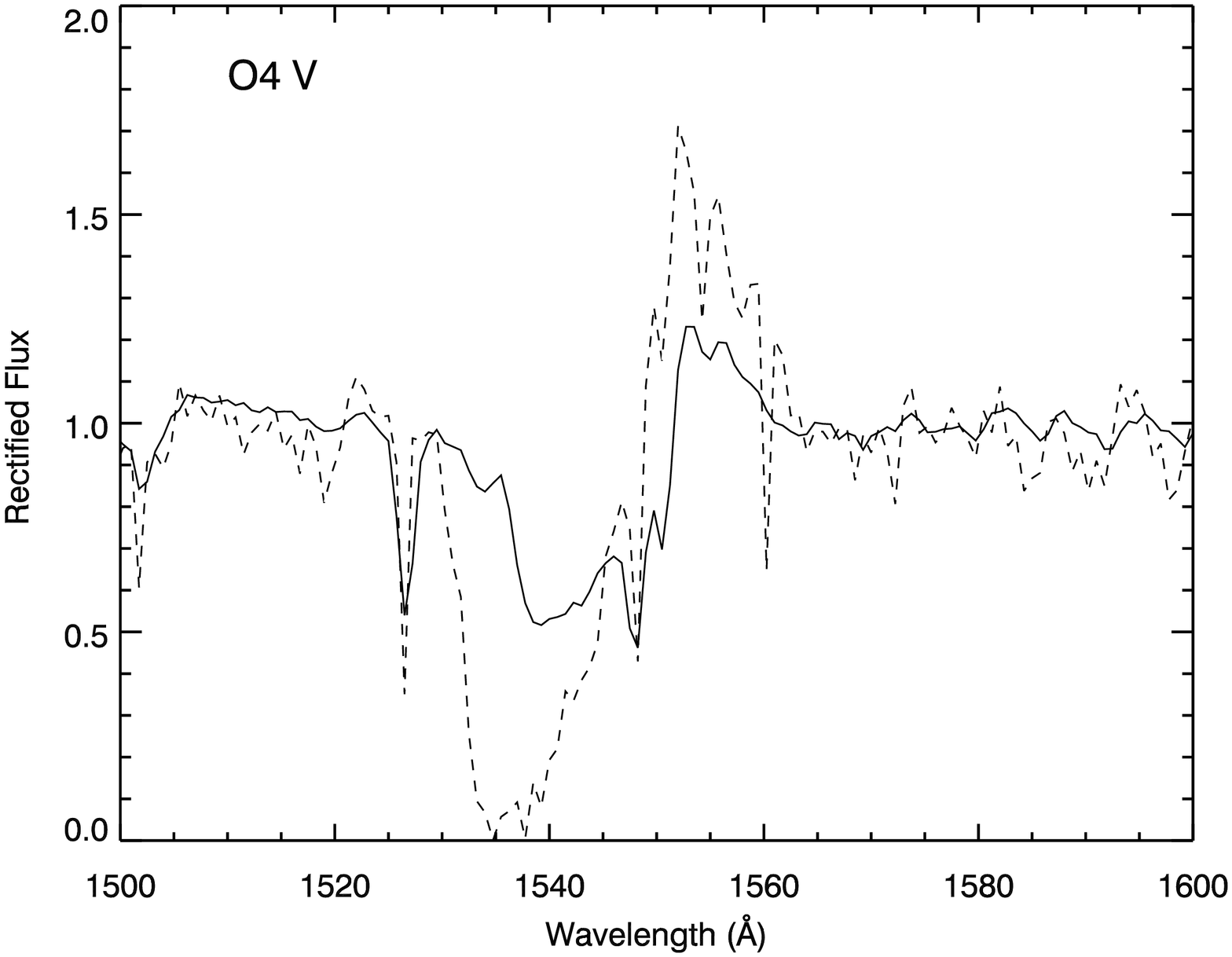,angle=90,width=9cm,height=6.5cm}
\medskip
Fig.~\ref{o4v}
\clearpage

\psfig{figure=o9v_nv.ps,angle=90,width=9cm,height=6.5cm}
\medskip
\psfig{figure=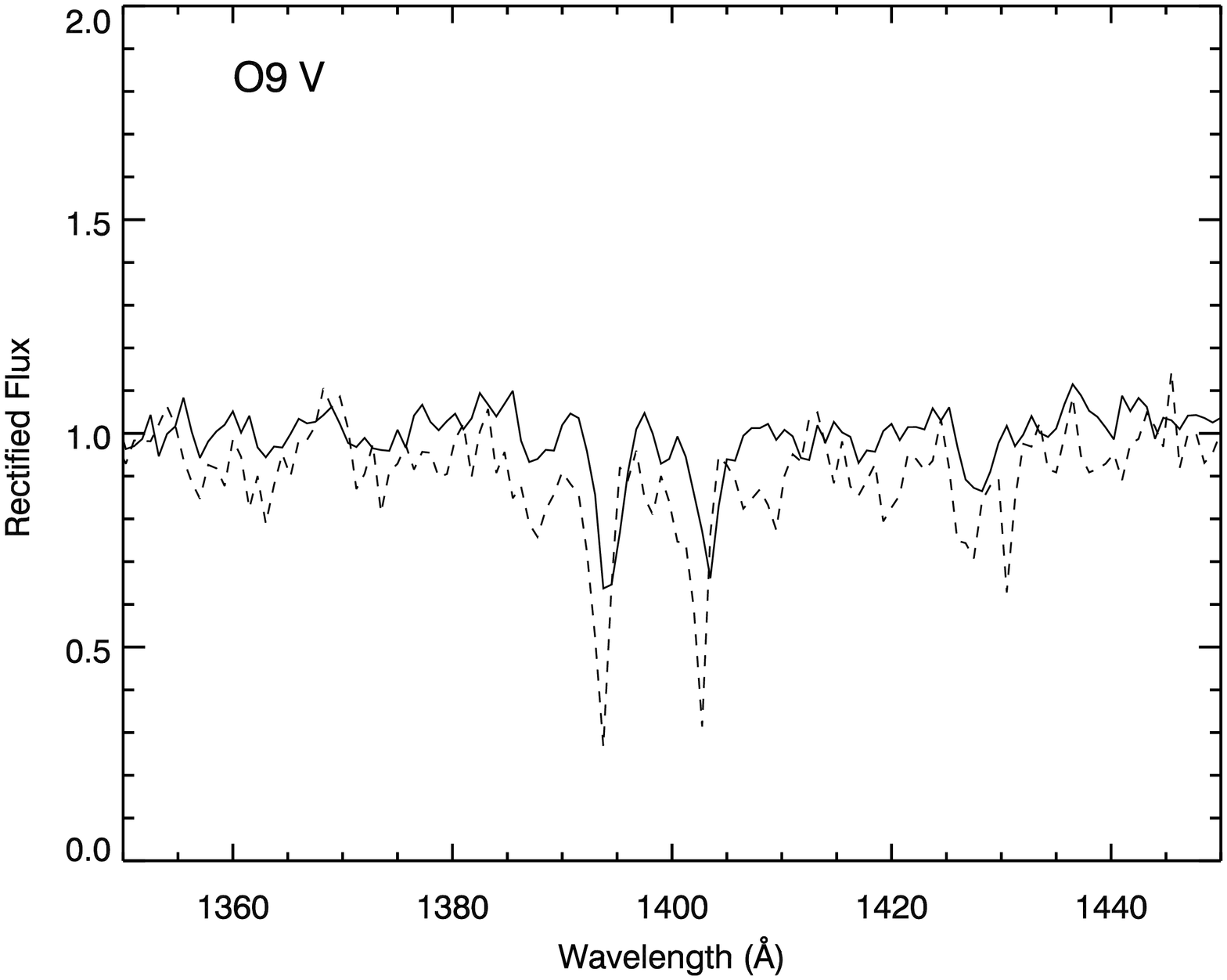,angle=90,width=9cm,height=6.5cm}
\medskip
\psfig{figure=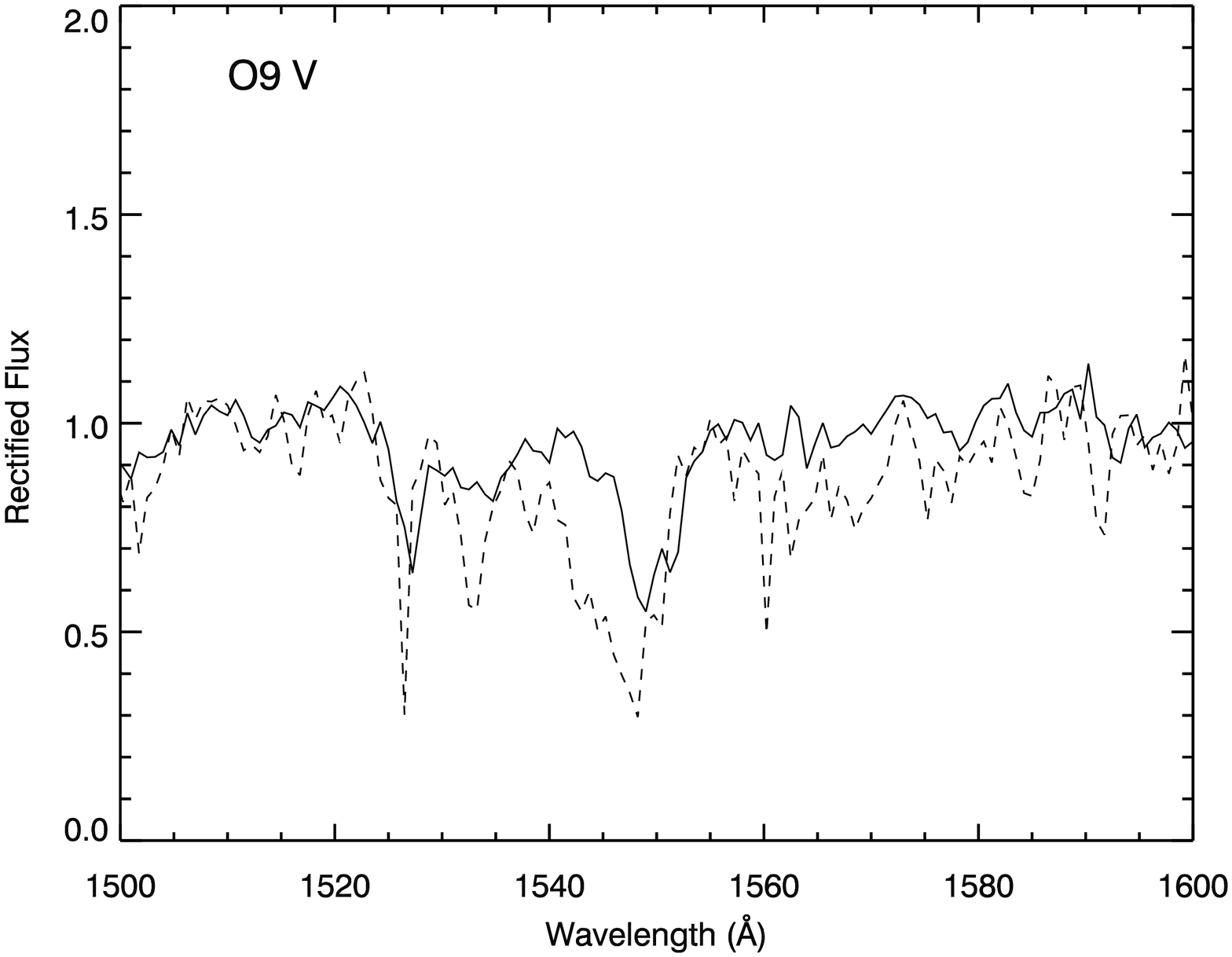,angle=90,width=9cm,height=6.5cm}
\medskip
Fig.~\ref{o9v}
\clearpage

\psfig{figure=o4i_nv.ps,angle=90,width=9cm,height=6.5cm}
\medskip
\psfig{figure=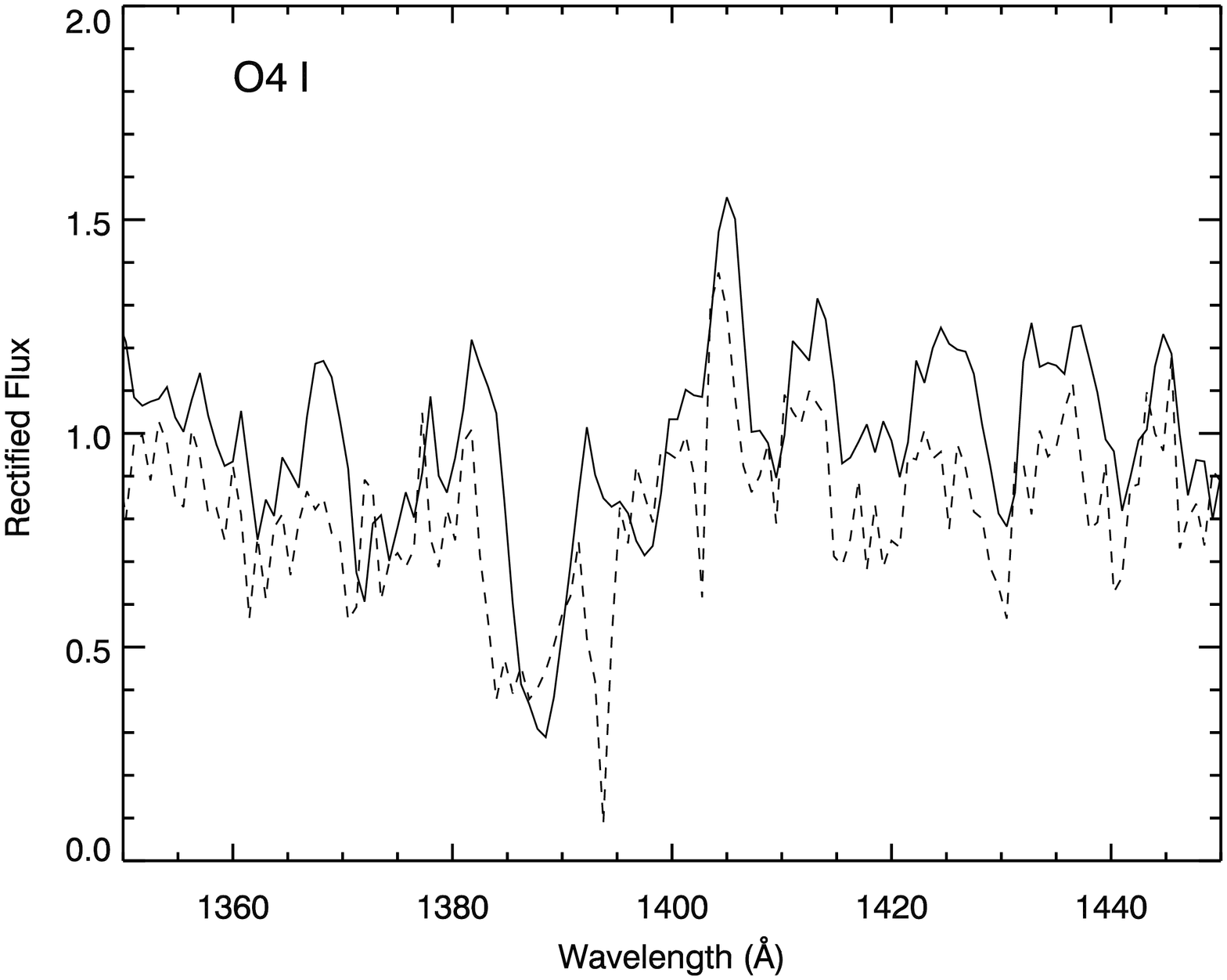,angle=90,width=9cm,height=6.5cm}
\medskip
\psfig{figure=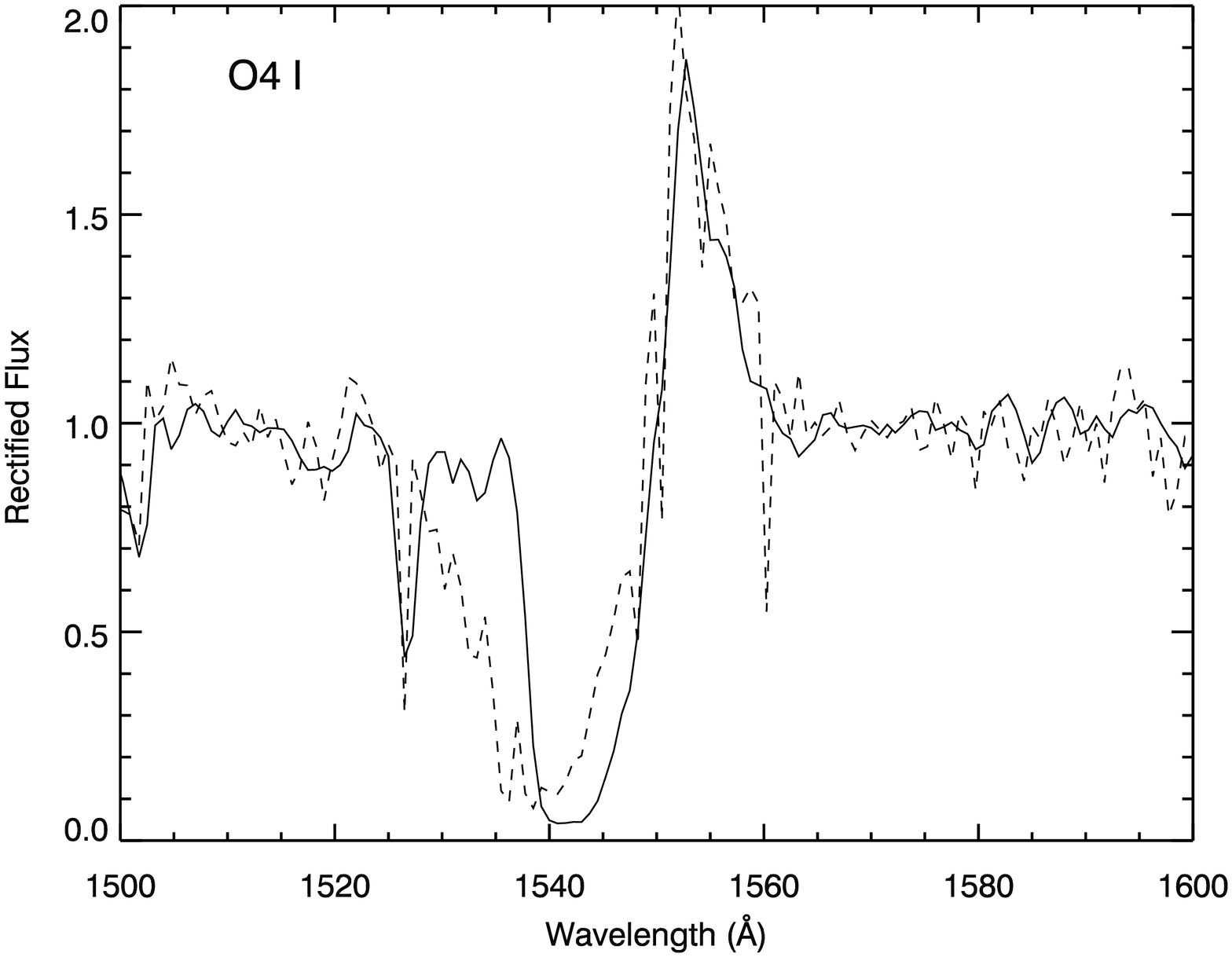,angle=90,width=9cm,height=6.5cm}
\medskip
Fig.~\ref{o4i}
\clearpage

\psfig{figure=o9i_nv.ps,angle=90,width=9cm,height=6.5cm}
\medskip
\psfig{figure=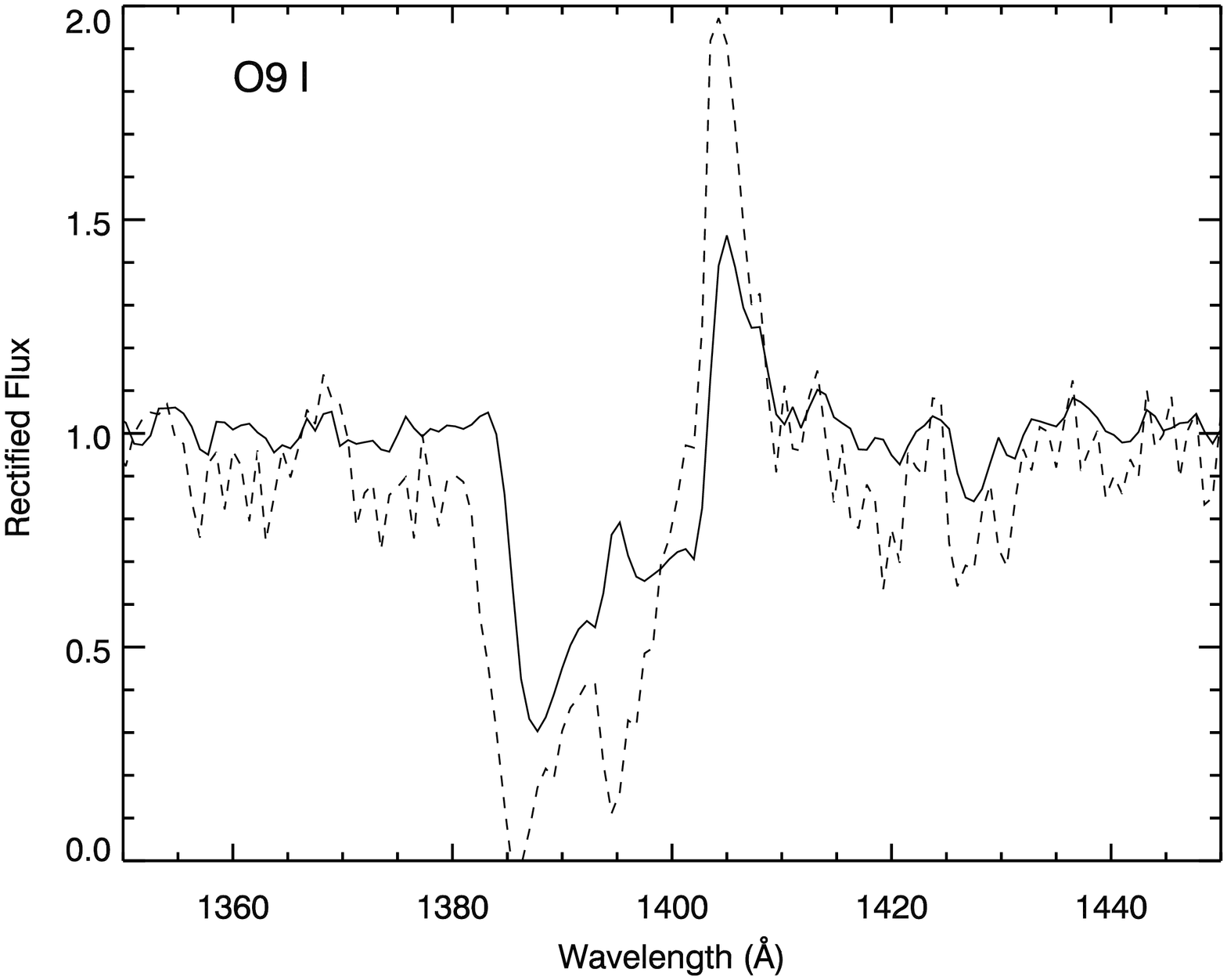,angle=90,width=9cm,height=6.5cm}
\medskip
\psfig{figure=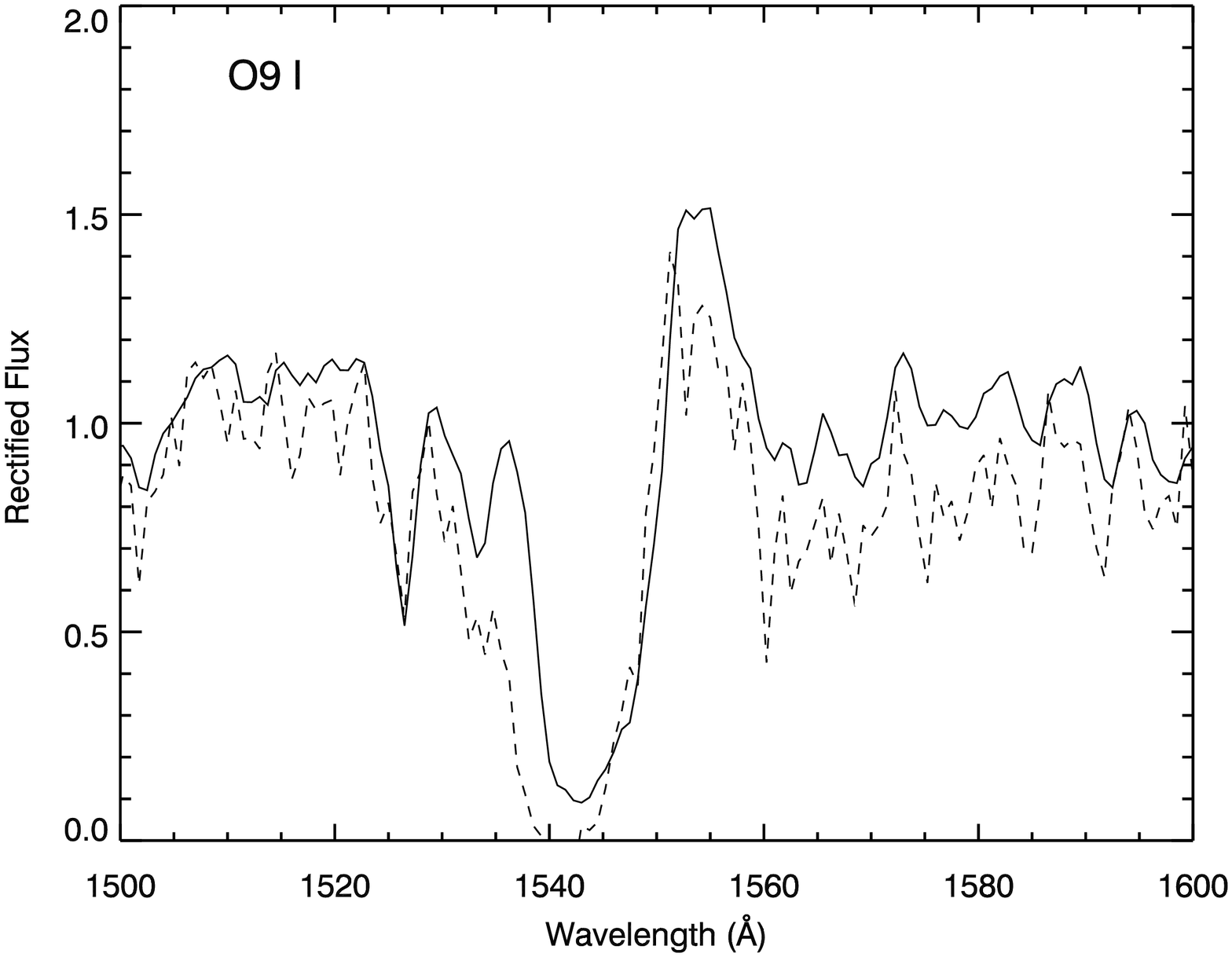,angle=90,width=9cm,height=6.5cm}
\medskip
Fig.~\ref{o9i}

\clearpage
\psfig{figure=zero_cont.ps,angle=180,width=15cm,height=20cm}
\medskip
Fig.~\ref{zero}

\clearpage

\psfig{figure=inst_stack.ps,angle=90,width=17cm,height=14cm}
Fig.~\ref{inst}

\clearpage
\psfig{figure=comp_0myr.ps,angle=90,width=13cm,height=9cm}
\vspace{-4.25cm}
\psfig{figure=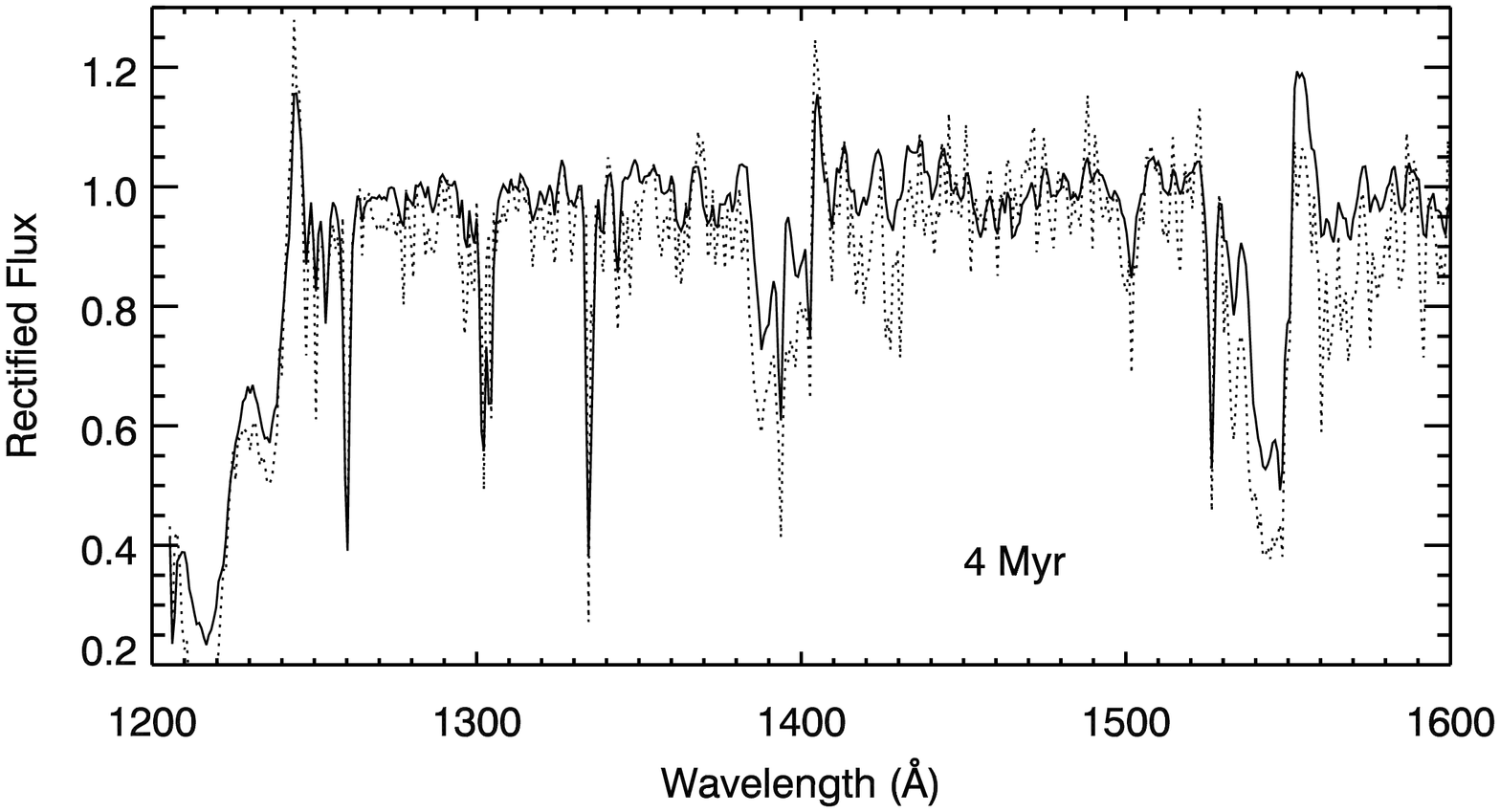,angle=90,width=13cm,height=9cm}
\vspace{-2.4cm}
\psfig{figure=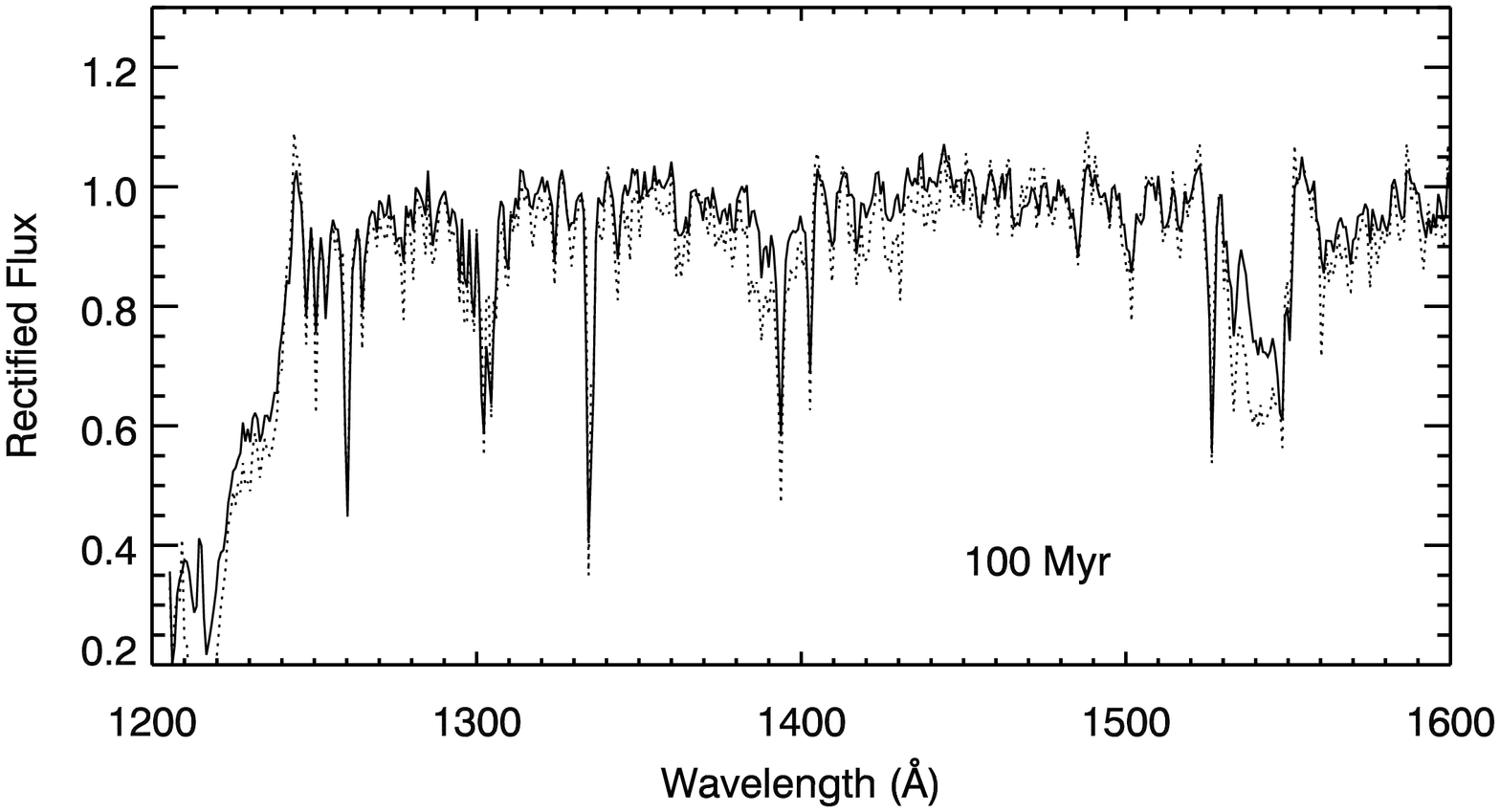,angle=90,width=13cm,height=9cm}
\vspace{-1cm}
Fig.~\ref{comp}

\clearpage
\psfig{figure=index.ps,angle=90,width=17cm,height=14cm}
Fig.~\ref{index}

\clearpage
\psfig{figure=cont_stack.ps,angle=90,width=17cm,height=14cm}
Fig.~\ref{cont}
\clearpage

\psfig{figure=n5253_comp.ps,angle=180,width=15cm,height=20cm}
Fig.~\ref{n5253}
\clearpage

\psfig{figure=abundance.ps,angle=90,width=17cm,height=14cm}
Fig.~\ref{abundances}
\clearpage

\psfig{figure=cB58.ps,angle=180,width=15cm,height=20cm}
Fig.~\ref{cB58}
\clearpage


\clearpage

\begin{table}
\caption{Library stars}
\scriptsize
\begin{center}
\begin{tabular}{l c c c c c}
\hline
\hline
\multicolumn{1}{l}{Identifier} &
\multicolumn{1}{c}{Galaxy} &
\multicolumn{1}{c}{Spectral Type} &
\multicolumn{1}{c}{$E(B-V)$} &
\multicolumn{1}{c}{Instrument} &
\multicolumn{1}{c}{HST Program} \\
\hline
NGC 346\#355        &  SMC  & ON3 III(f$^*$) & 0.08 &  FOS,STIS & 4110,7437\\
Sk--66$^\circ$172   &  LMC  &  O3 III(f$^*$) & 0.19 &  FOS & 2233\\
Sk--68$^\circ$137   &  LMC  &  O3 III(f$^*$) & 0.24 &  FOS & 2233\\
AV 14               &  SMC  &  O3--4 V       & 0.12 &  FOS & 5444\\
AV 388              &  SMC  &  O4 V          & 0.10 &  FOS & 4110\\
NGC 346\#324        &  SMC  &  O4 V((f))     & 0.07 &  FOS,STIS & 4110,7437\\
Sk--70$^\circ$60    &  LMC  &  O4 V          & 0.12 &  FOS & 5444\\
NGC 346\#1          &  SMC  &  O4 III(n)(f)  & 0.11 &  FOS & 4110\\
Sk--67$^\circ$166   &  LMC  &  O4 If+        & 0.09 &  FOS & 4110\\
NGC 346\#368        &  SMC  &  O4--5 V((f))  & 0.08 &  STIS& 7437\\
AV 80               &  SMC  &  O4--6n(f)p    & 0.17 &  STIS& 7437\\
AV 61               &  SMC  &  O5 V          & 0.07 &  FOS & 5444\\
Sk--70$^\circ$69    &  LMC  &  O5 V          & 0.06 &  FOS & 2233,5444\\
AV 75               &  SMC  &  O5 III(f+)    & 0.15 &  FOS,STIS & 5444,7437\\
NGC 346\#4          &  SMC  &  O5--6 V       & 0.08 &  FOS & 4110\\
AV 243              &  SMC  &  O6 V          & 0.09 &  FOS & 4110\\
AV 377              &  SMC  &  O6 V          & 0.00 &  FOS & 5444\\
NGC 346\#113        &  SMC  & OC6 V          & 0.09 &  STIS& 7437\\
Sk--66$^\circ$100   &  LMC  &  O6 II(f)      & 0.10 &  FOS & 2233\\
HDE 269357          &  LMC  &  O6 I          & 0.10 &  FOS & 5444\\
AV 220              &  SMC  &  O6.5f?p       & 0.09 &  STIS &7437\\
Sk--70$^\circ$91    &  LMC  &  O6.5 V        & 0.08 &  FOS & 5444\\
AV 15               &  SMC  &  O6.5 II(f)    & 0.09 &  FOS,STIS& 5444,7437\\
AV 207              &  SMC  &  O7 V          & 0.09 &  FOS & 5444\\
BI 155              &  LMC  &  O7 V          & 0.11 &  FOS & 5444\\
BI 208              &  LMC  &  O7 V          & 0.02 &  FOS & 5444\\
AV 26               &  SMC  &  O7 III        & 0.13 &  FOS & 5444\\
AV 95               &  SMC  &  O7 III((f))   & 0.01 &  STIS& 7437\\
BI 229              &  LMC  &  O7 III        & 0.14 &  FOS & 5444\\
BI 272              &  LMC  &  O7 II         & 0.16 &  FOS & 5444\\
AV 83               &  SMC  &  O7 Iaf+       & 0.18 &  STIS &7437\\
AV 232              &  SMC  &  O7 Iaf+       & 0.10 &  FOS & 4110\\
AV 69               &  SMC  & OC7.5 III((f)) & 0.09 &  STIS &7437\\
AV 378              &  SMC  &  O8 V          & 0.07 &  FOS & 5444\\
Sk--67$^\circ$191   &  LMC  &  O8 V          & 0.10 &  FOS & 5444\\
AV 47               &  SMC  &  O8 III((f))   & 0.05 &  FOS,STIS & 5444,7437\\
BI 9                &  LMC  &  O8 III        & 0.14 &  FOS & 5444\\
BI 173              &  LMC  &  O8 III        & 0.17 &  FOS & 5444\\
AV 469              &  SMC  &  O8 II         & 0.09 &  FOS & 5444\\
AV 396              &  SMC  &  O9 V          & 0.08 &  FOS & 5444\\
AV 451              &  SMC  &  O9 V          & 0.08 &  FOS & 5444\\
AV 223              &  SMC  &  O9 III        & 0.11 &  FOS & 5444\\
AV 238              &  SMC  &  O9 III        & 0.09 &  FOS & 4110\\
Sk--67$^\circ$101   &  LMC  &  O9 III        & 0.14 &  FOS & 5444\\
Sk--69$^\circ$124   &  LMC  &  O9 Ib         & 0.12 &  FOS & 5444\\
AV 372              &  SMC  &  O9 I          & 0.13 &  FOS & 5444\\
BI 192              &  LMC  &  O9.5 III      & 0.12 &  FOS & 5444\\
BI 170              &  LMC  &  O9.5 II       & 0.14 &  FOS & 5444\\
AV 327              &  SMC  &  O9.5 II--Ibw  & 0.09 &  STIS &7437\\
AV 170              &  SMC  &  O9.7 III      & 0.08 &  STIS &7437\\
Sk--65$^\circ$21    &  LMC  &  O9.7 Iab      & 0.15 &  FOS & 4110\\
NGC 346\#12         &  SMC  &  O9.5--B0 V    & 0.16 &  STIS& 7437\\
AV 488              &  SMC  &  B0.5 Iaw      & 0.18 &  FOS & 4110\\
\hline
\end{tabular}
\end{center}
\end{table}

\end{document}